\documentclass[aps,prb,twocolumn,floatfix,10pt]{revtex4-1}
\usepackage[utf8]{inputenc}
\usepackage{natbib}
\usepackage{graphicx}
\usepackage{latexsym}
\usepackage{amssymb}
\usepackage{mathtools}
\usepackage{amsfonts}
\usepackage{gensymb}
\usepackage{bm}
\usepackage{hyperref}
\usepackage{braket}
\usepackage{physics}
\usepackage{bbold}
\usepackage[caption=false]{subfig}
\usepackage[dvipsnames]{xcolor}
\usepackage[normalem]{ulem}

\begin{document}

\title{Fractonic order and emergent fermionic gauge theory}
\author{Wilbur Shirley}
\affiliation{Department of Physics and Institute for Quantum Information and Matter, California Institute of Technology, Pasadena, California 91125, USA}

\date{\today}

\begin{abstract}

We consider fermionic systems in which fermion parity is conserved within rigid subsystems, and describe an explicit procedure for gauging such subsystem fermion parity symmetries to obtain bosonic spin Hamiltonians. We show that gauging planar or fractal subsystem fermion parity symmetry in three spatial dimensions gives rise to a plethora of exactly solvable spin models exhibiting novel gapped fractonic orders characterized by emergent fermionic gauge theory. The low energy excitations of these models include fractional quasiparticles with constrained mobility and emergent fermionic  statistics. We illustrate this phenomenon through a series of examples including fermionic analogs of both foliated fracton phases and fractal spin liquids. We find that the foliated analogs actually exhibit the same fractonic order as their bosonic counterparts, while this is not generally the case for fermionic fractal spin liquids.

\end{abstract}

\maketitle

\section{Introduction}

Gauge theory is a unifying framework in the study of gapped quantum matter. Many of the paradigmatic examples of intrinsic topological order\cite{WenRigid} in two spatial dimensions, such as the Kitaev toric code and quantum double models,\cite{KitaevToricCode} are characterized by emergent gauge theory with a discrete gauge group $G$.\cite{DijkgraafWitten} In three spatial dimensions, it has been conjectured that all intrinsic topological orders are characterized by emergent gauge theory of either a bosonic or fermionic nature.\cite{Lan1,Lan2}
In systems with global symmetry, gauging the symmetry by enhancing it to a local symmetry can be a powerful tool to probe the properties of the system. For instance, 2D symmetry-protected topological (SPT) phases\cite{Xie13} with internal symmetry are characterized by the nontrivial braiding and exchange statistics of gauge flux excitations in the `twisted' gauge theories obtained by gauging the symmetry.\cite{LevinGuSPT}

A highly fruitful generalization of the gauging procedure has been in systems that have \textit{subsystem} symmetries---internal symmetries that act on rigid subsystems of a many-body lattice model. It is now well-understood that gauging these kinds of symmetry gives rise an exotic new class of 3D gapped quantum matter said to exhibit \textit{fractonic order}.\cite{FractonRev,VijayFracton,Sagar16,ChamonModel,HaahCode,YoshidaFractal,Slagle17Lattices,CageNet,VijayNonabelian,MaLayers,ChamonModel2,HsiehPartons,HalaszSpinChains,Slagle2spin,BulmashFractal,BulmashGauging,GaugingPermutation,YouLitinskiOppen,TQFT,NatSagar,XieReview,SongTwistedFracton,Zohar1,Zohar2,Gromov_2019,Robust} Fractonic order is a form of long-range entanglement in gapped lattice systems that is sensitive to the geometry of the underlying lattice.\cite{ChamonModel,ChamonModel2,HaahCode,YoshidaFractal,VijayFracton,Sagar16} This sensitivity manifests in a number of striking properties that collectively preclude a low energy topological quantum field theory (TQFT) description---most notably, 1) fractional excitations with fundamentally constrained mobility,\cite{ChamonModel,HaahCode,YoshidaFractal,VijayFracton} 2) a robust ground state degeneracy on a torus that grows exponentially with length of the system,\cite{ChamonModel,HaahCode,VijayFracton} and 3) linear subleading corrections to the ground state area law of entanglement.\cite{SwingleSSource,ShiEntropy,HermeleEntropy,BernevigEntropy,FractonEntanglement} The constraints on quasiparticle mobility come in a variety of forms---the excitations in fractonic phases of matter can be roughly categorized as \textit{planons}, \textit{lineons}, and \textit{fractons}, which are respectively mobile within a plane, mobile along a line, and fully immobile as individual particles. Broadly speaking, fractonic orders may be divided into `planar' (for instance the X-cube model)\cite{Sagar16} and `fractal' orders (for example the Haah code)\cite{HaahCode} according to the absence or presence of fractal-like fusion rules amongst these excitations.

Thus far, subsystem symmetries have been studied in the setting of bosonic spin systems. In this paper, we consider \textit{subsystem fermion parity symmetries}, which measure the fermion parity within rigid subsystems of a fermionic lattice model. We describe an explicit procedure for gauging such symmetries, under which physical fermions are mapped into emergent fermionic gauge charges in a spin model without any microscopic fermion degrees of freedom. This procedure is analogous to gauging the \textit{global} fermion parity symmetry in fermionic systems in 2D or 3D to obtain fermionic gauge theories, for instance the phases of the Kitaev 16-fold way\cite{KitaevAnyons} in 2D and the fermionic toric code in 3D.\cite{LevinWen,WalkerWang} On a technical level, the gauging procedure we describe is similar in spirit to the `bosonization' dualities introduced in Refs. \onlinecite{YuanChen1,YuanChen2,YuanChen3} for fermionic systems in two, three, and higher dimensions. It can be viewed as a bosonization of the subsystem symmetric fermionic operator algebra.

Whereas topological gauge theories contain fully deconfined fermionic gauge charge, the fractonic gauge theories introduced in this paper exhibit gauge charge with constrained mobility as a consequence of conservation of subsystem symmetry charge in the ungauged system.\cite{GaugingSubsystem} We demonstrate that the exchange statistics of charged lineons and planons---well-defined universal properties that characterize a given fractonic order---are inherited from the statistics of the degrees of freedom in the ungauged model. Although the notion of statistics for fractons is much more subtle because they are immobile particles and thus cannot be exchanged, we argue that driving a condensation of fracton dipoles reveals the fermionicity of fractons in a fermionic analog of the X-cube model. We illustrate this correspondence in both planar and fractal fractonic orders, collectively exhibiting examples of fermionic fractons, lineons, and planons. Surprisingly, we find that in the case of planar fractonic orders, it is possible to transmute the statistics of gauge charges from fermionic to bosonic or vice versa via a flux attachment procedure. As a result there is an equivalence between the fermionic and bosonic gauge theories. This is not the case in general for fermionic fractal spin liquids, which represent unique fractonic orders from their bosonic counterparts.

The paper is organized as follows. In Section \ref{sec:global}, we review the procedure of gauging global fermion parity to obtain topologically ordered spin Hamiltonians in 2D and 3D. In Section \ref{sec:subsystem}, we describe a general procedure for gauging subsystem fermion parity in translation-invariant fermionic systems to obtain bosonic spin models with fractonic order, and discuss a handful of illustrative examples in Section \ref{sec:examples}. In Section \ref{sec:statistics} we describe statistical processes which capture the fermionic nature of the lineon and planon gauge charge excitations in these models. In Section \ref{sec:unique} we address the question of whether these models represent fractonic orders unique from their bosonic gauge theory counterparts. We conclude with a discussion in Section \ref{sec:discussion}.

\section{Gauging global fermion parity}

\label{sec:global}

First, we will review the procedure of gauging global fermion parity symmetry in translation-invariant systems in both two and three spatial dimensions. Gauging fermion parity in a trivial 2D insulating state gives rise to 2D toric code topological order,\cite{KitaevToricCode} mapping physical fermion excitations to the emergent fermionic gauge charge $\epsilon$. In 3D, physical fermions are mapped under gauging to fermionic gauge charges of the fermionic 3D toric code, which is a topologically ordered spin model distinct from the ordinary 3D toric code.\cite{LevinWen,WalkerWang} The procedures we describe utilize operator algebra mappings which are similar in spirit to the bosonization dualities introduced in Refs. \onlinecite{YuanChen1,YuanChen2,YuanChen3}. However, the dualities we employ take a slightly different form which is more amenable to generalization to subsystem symmetries.

\subsection{Two dimensions}

We begin with a trivial insulating state in 2D. The system contains a spinless fermionic mode at each site $i$ of a square lattice, described by fermion creation and annihilation operators $a_i^\dagger$ and $a_i$ satisfying the usual anticommutation relations. We will use Majorana operators $\gamma_i=a_i+a^\dagger_i$ and $\gamma'_i=(a_i-a^\dagger_i)/i$ which satisfy $\{\gamma_i,\gamma_j\}=\{\gamma'_i,\gamma'_j\}=2\delta_{ij}$ and $\{\gamma_i,\gamma'_j\}=0$. The Hamiltonian takes the form
\begin{equation}
    H_0=i\sum_i\gamma_i\gamma'_i=\sum_i\left(2a_i^\dagger a_i-1\right),
    \label{eq:H}
\end{equation}
whose ground state is simply the state with zero occupation number on every site. Although the gauging procedure can be implemented on any local fermionic Hamiltonian, we focus on the trivial Hamiltonian $H_0$ because the gauged model it yields is an exactly solvable, zero-correlation length model (in fact, it is equivalent to the 2D toric code).

The global fermion parity symmetry is a unitary operator $P$ that acts as $(-1)^{F}$ where $F$ is the total fermion number, in particular
\begin{equation}
    P=\prod_i-i\gamma_i\gamma'_i.
\end{equation}
Note that $P^2=1$ hence fermion parity is a `$Z_2$' symmetry denoted $Z_2^f$. To gauge the fermion parity symmetry, we first expand the Hilbert space by introducing $Z_2$ gauge fields, i.e. qubit degrees of freedom, on each link $ij$ of the lattice, and furthermore enforce the Gauss's law
\begin{equation}
    -i\gamma_i\gamma'_i\prod_j X_{ij}=1
    \label{eq:constraint}
\end{equation}
as a hard constraint on the Hilbert space. (Here $X_{ij}$ refers to the Pauli $X$ operator on link $ij$, and $j$ indexes all nearest neighbors of site $i$). The Gauss's law equates the fermion number at each site to the divergence of the $Z_2$ `electric field'. The algebra of fermion parity-preserving operators is generated by the on-site parity operators $-i\gamma_i\gamma'_i$ and the two-site quadratic `minimal couplings' $\gamma_i\gamma_j$. Thus, to couple the fermionic Hamiltonian to the gauge field, we replace the minimal couplings $\gamma_i\gamma_j$ with the gauge-invariant (i.e. Gauss's law-preserving) operators $\gamma_iZ_{ij}\gamma_j$ throughout, where $Z_{ij}$ is the Pauli $Z$ operator on link $ij$. In the case of the trivial Hamiltonian $H_0$ this step does nothing but is necessary in general. The next step is to add plaquette operators $B_p=Z_{ij}Z_{jk}Z_{kl}Z_{li}$ to the Hamiltonian to gap out the gauge flux excitations, yielding the gauged Hamiltonian
\begin{equation}
    {H}_g=i\sum_i\gamma_i\gamma'_i -\sum_p B_p.
\end{equation}

\begin{figure}
    \centering
    \includegraphics[width=0.4\textwidth]{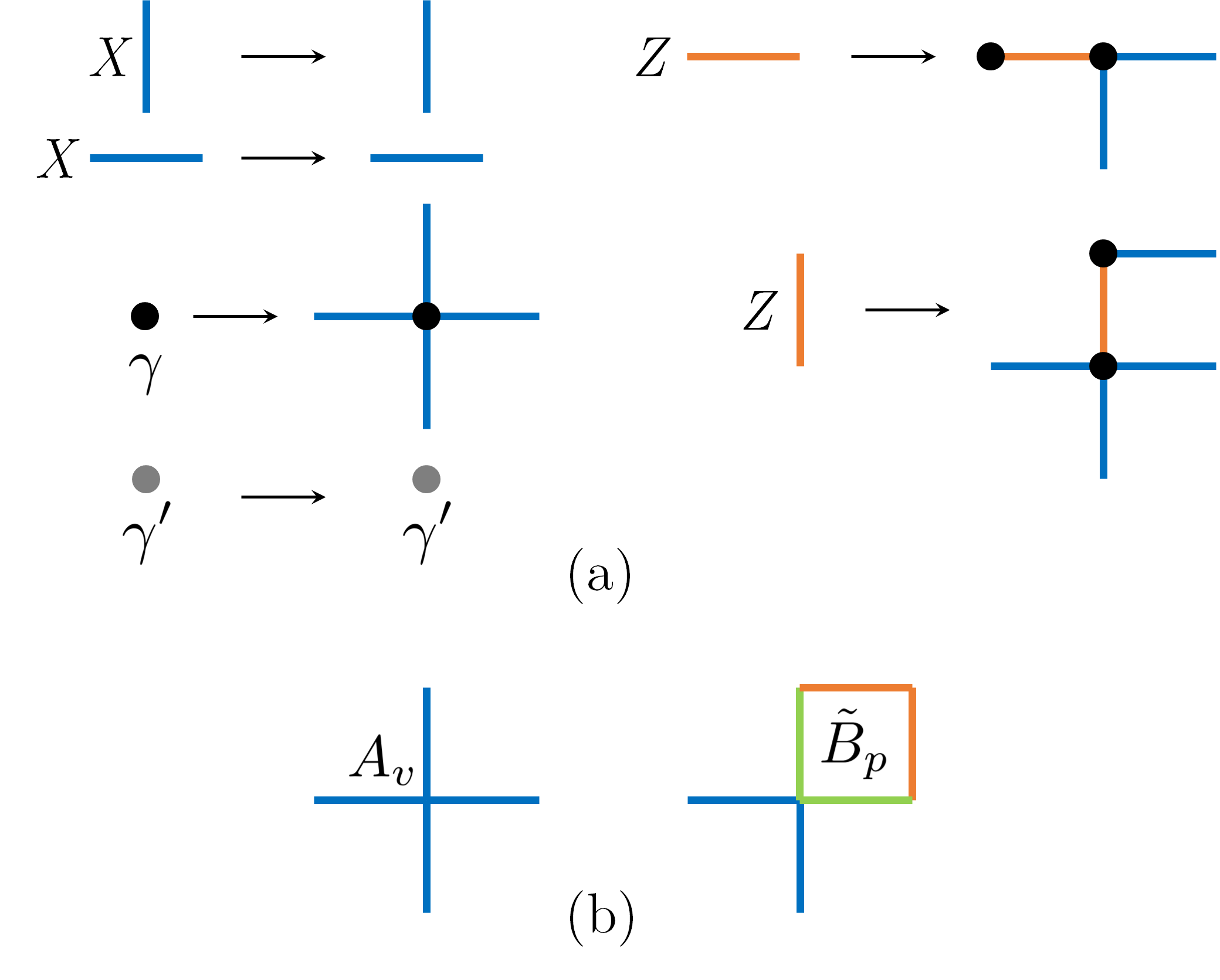}
    \caption{(a) Action of the local unitary circuit that maps $H_g$ to $H_\text{2D}$. Here, blue and orange represent Pauli $X$ and $Z$ operators on the link degrees of freedom respectively, while black and gray dots represent the $\gamma$ and $\gamma'$ Majorana operators on a particular site. This mapping preserves all commutation and anticommutation relations of the operator algebra. (b) The vertex and plaquette terms $A_v$ and $\tilde{B}_p$ of the Hamiltonian $H_2D$. They are tensor products of Pauli operators. Here, light green represents Pauli $Y$.}
    \label{fig:2Dgaugecv}
\end{figure}

The final step is to transform the Hamiltonian via a local unitary operator that maps the Gauss's law constraint (\ref{eq:constraint}) to the new constraint $-i\gamma_i\gamma'_i=1$ on the fermionic parity at each lattice site $i$. This constraint suppresses the fermionic degrees of freedom and results in a purely bosonic spin model with a tensor product Hilbert space. There is not a unique choice of local unitary that will accomplish this task; one particular choice is depicted pictorially in Fig. \ref{fig:2Dgaugecv}a, which defines the local unitary by its action on the operator algebra.
After an additional local unitary $Y\leftrightarrow Z$ on each gauge qubit the resulting spin Hamiltonian takes the form
\begin{equation}
    H_\text{2D}=-\sum_v A_v-\sum_p \tilde{B}_p
\end{equation}
where $A_v$ and $\tilde{B}_p$ are depicted in Fig. \ref{fig:2Dgaugecv}b. This Hamiltonian bears similarity to the 2D toric code,\cite{KitaevToricCode} but differs in that the plaquette term $\tilde{B}_p$ is a product of the vertex and plaquette terms $A_v$ and $B_p$ of the original toric code.

Altogether, the gauging procedure maps physical fermionic excitations of the ungauged insulator to excitations of the emergent gauge symmetry $A_v$, which are fermionic gauge charges belonging to the $\epsilon$ superselection sector of the 2D toric code. The $\epsilon$ fermion is a bound state of $e$ and $m$ bosons, which are interpreted as gauge fluxes.  Note that the product of $\tilde{B}_p$ operators in a large region is equal to a loop operator corresponding to a process of creation, movement, and annihilation of a pair of emergent fermions.

\subsection{Three dimensions}

\begin{figure}
    \centering
    \includegraphics[width=0.4\textwidth]{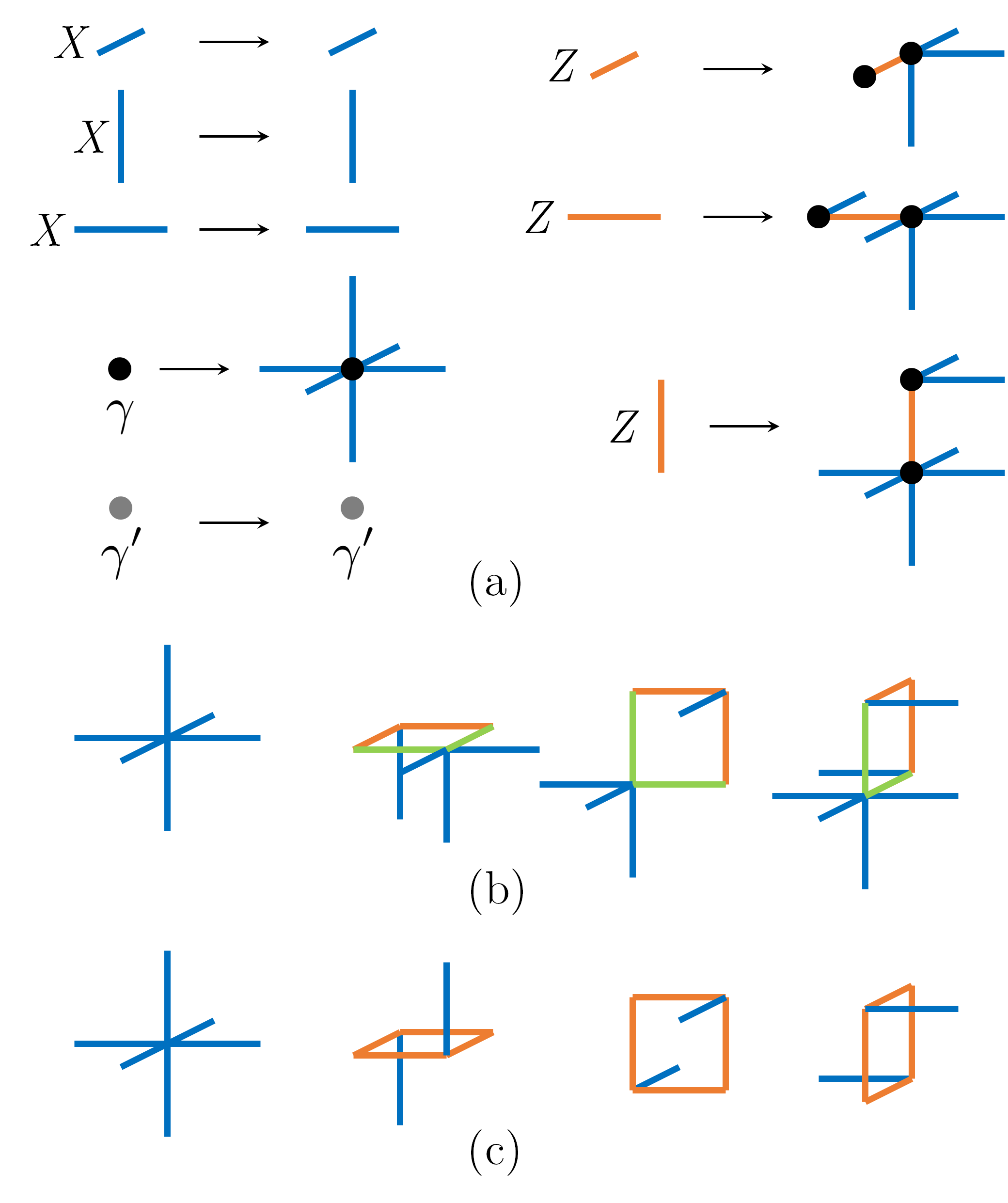}
    \caption{(a) Action of the circuit that maps $H_g$ to $H_\text{3D}$. (b) Vertex and plaquette terms $A_v$ and $\tilde{B}_p$ of the Hamiltonian $H_\text{3D}$, which are tensor products of Pauli operators. In this figure, blue, light green, and orange represent Pauli $X$, $Y$, and $Z$ respectively, while black and gray dots represent the $\gamma$ and $\gamma'$ Majorana operators on a particular site. (c) The Walker-Wang Hamiltonian for the fermionic 3D toric code, which can be obtained from $H_\text{3D}$ by composing each plaquette operator with an appropriate vertex term.}
    \label{fig:3Dgaugecv}
\end{figure}

Now let us consider a trivial insulator on a cubic lattice with one fermionic mode per site. The Hamiltonian takes the same form $H_0$. The gauging procedure is very similar to the 2D case. First, we add a $Z_2$ gauge field to each link of the lattice, and enforce the Gauss's law constraint (\ref{eq:constraint}). In general the Hamiltonian is coupled to the gauge field by replacing the two-body minimal couplings $\gamma_i\gamma_j$ with the gauge-invariant operators $\gamma_iZ_{ij}\gamma_j$. The gauged Hamiltonian once again takes the form 
\begin{equation}
    {H}_g=i\sum_i\gamma_i\gamma'_i -\sum_p B_p
\end{equation}
To eliminate the fermionic degrees of freedom we consider the change of variables depicted in Fig. \ref{fig:3Dgaugecv}a, which maps the Gauss's law constraint (\ref{eq:constraint}) to the on-site fermion parity constraint $-i\gamma_i\gamma'_i=1$. After the additional unitary $Y\leftrightarrow Z$ on each link the resulting spin Hamiltonian takes the form
\begin{equation}
    H_\text{3D}=-\sum_v A_v-\sum_p \tilde{B}_p
\end{equation}
where $A_v$ and $B_p$ are depicted in Fig. \ref{fig:3Dgaugecv}b. By multiplying each plaquette term by an appropriate vertex term, the Hamiltonian can be transformed into the familiar Walker-Wang form\cite{WalkerWang} shown in Fig. \ref{fig:3Dgaugecv}c.

As in the 2D case, the vertex term $A_v$ is interpreted as an emergent gauge symmetry whose excitations are fermionic gauge charges corresponding to the physical fermions in the ungauged insulator. For this reason these models represent a distinct topological order from the ordinary 3D toric code.\cite{LevinWen} The product of $\tilde{B}_p$ operators over a large membrane is equal to a loop operator corresponding to a process of creation, movement, and annihilation of a pair of fermionic charges.

\section{Gauging subsystem fermion parity symmetry}

\label{sec:subsystem}

In this section we describe a general procedure for gauging subsystem fermion parity symmetry in fermionic lattice systems. It generalizes the well-known methods of gauging both subsystem symmetries in bosonic systems\cite{Sagar16,WilliamsonUngauging,KubicaYoshidaUngauging,GaugingSubsystem} and global fermion parity symmetry in fermionic systems. Conceptually, the idea is to introduce auxiliary gauge fields which mediate the symmetry-preserving interactions between fermions. These gauge fields are then coupled with the original matter degrees of freedom by generalized Gauss's laws which relate the fermion density to the configuration of the gauge field. The main technical subtlety arises in transforming the gauged Hamiltonian so that it acts on a purely bosonic Hilbert space; these transformations amount to a `bosonization' of the subsystem fermion parity-preserving operator algebra.

The input to the procedure is a fermionic lattice model, a set of subsystem fermion parity symmetries to be gauged, and a local Hamiltonian preserving these symmetries. A subsystem fermion parity symmetry $P_S$ acts as $(-1)^{F_S}$ where $F_S$ is the total fermion number within a particular subsystem $S$. In other words,
\begin{equation}
    P_S=\prod_{i\in S}-i\gamma_i\gamma'_i.
\end{equation}
For simplicity we will assume that the ungauged model is defined on a Bravais lattice with a single fermionic mode per site. However, the procedure is readily generalized to systems with additional degrees of freedom including bosonic ones; we will discuss one such example in Section \ref{sec:half}. We will also assume that the global fermion parity is generated by the subsystem fermion parity symmetries, so that the system becomes bosonic upon gauging. It is also possible to gauge subsystem symmetry groups that do not include global fermion parity, however in this case the gauged system will retain local fermionic degrees of freedom.

The gauging procedure can be summarized as follows:
\begin{enumerate}
    \item Identify a set of multi-body `minimal coupling' terms $\{c_\alpha\}$ that, together with the on-site fermion parities $-i\gamma_i\gamma'_i$, generate the algebra of subsystem fermion parity-preserving operators. It is always possible to choose couplings of the form $c_\alpha=\prod_{i\in\alpha}\gamma_i$ which do not depend on $\{\gamma'_i\}$. Here we abuse notation to allow $\alpha$ to index the couplings and simultaneously denote the support of $c_\alpha$. Note that $c_\alpha$ must be even-body in order to preserve global fermion parity.
    \item Expand the Hilbert space by introducing a $Z_2$ gauge degree of freedom $\sigma_\alpha$ corresponding to each minimal coupling $c_\alpha$. Subsequently restrict the Hilbert space by imposing hard Gauss's law constraints of the form $-i\gamma_i\gamma'_i\prod_{\alpha\ni i} X_\alpha=1$.
    \item Couple the Hamiltonian to the gauge fields by replacing minimal coupling terms $c_\alpha$  with the gauge-invariant terms $c_\alpha Z_\alpha$ throughout.
    \item Add terms to the Hamiltonian to gap out the gauge flux excitations. These may be chosen to be local tensor products of gauge field Pauli $Z$ operators that preserve the Gauss's law constraint.
    \item Use a local unitary circuit to map the Gauss's law constraint at site $i$ to the constraint $-i\gamma_i\gamma'_i=1$ on the fermion parity at $i$, thus suppressing the fermionic degrees of freedom and yielding a purely bosonic spin Hamiltonian. This unitary acts on the operator algebra as
        \begin{equation}
        \begin{split}
            X_\alpha&\to X_\alpha\qquad\qquad Z_\alpha\to c_\alpha Z_\alpha \prod_{\substack{\beta\\T(\alpha,\beta)=1}} X_\beta \\
            \gamma_i&\to \gamma_i\prod_{\alpha\ni i} X_\alpha\qquad\qquad \gamma_i'\to\gamma_i'
        \end{split}
        \label{eq:transform}
        \end{equation}
        where $T$ is an \textit{antisymmetric} function with the following property:
        \begin{equation}
            T(\alpha,\beta)=
            \begin{cases}
                \pm1 & \text{if } \{c_\alpha,c_\beta\}=0 \\
                0 & \text{if } [c_\alpha,c_\beta]=0
            \end{cases}
        \end{equation}
        It is readily verified that this mapping preserves the commutation and anticommutation relations of the operator algebra provided that $T$ satisfies this property.
        Intuitively, the action on $\gamma_i$ and $\gamma'_i$ is chosen so that the Gauss's law is mapped to the constraint $-i\gamma_i\gamma'_i=1$. To preserve the commutation between $Z_\alpha$ and $\gamma_i$, the operator $Z_\alpha$ must be composed with $c_\alpha$. Finally, $Z_\alpha$ must be composed with the additional factors of $X_\beta$ so that it commutes with the image of all $Z_\beta$.
        In order to produce a translation-invariant spin Hamiltonian, we will assume that $T$ is chosen to respect the translational symmetry. Note that even with this requirement there is not a unique choice of $T$.
    
\end{enumerate}

Steps 1 through 4 are identical in spirit to the procedure of gauging subsystem symmetries in bosonic systems. After these steps, the Hamiltonian acts on a fermionic Hilbert space constrained by the generalized Gauss's law, which is actually equivalent to a bosonic tensor product Hilbert space; step 5 transforms the Hamiltonian to make this equivalence manifest.

\section{Examples}
\label{sec:examples}

We will now illustrate this procedure with some concrete examples. Although the procedure can be applied to any fermionic Hamiltonian that preserves a given set of subsystem fermion parity symmetries, we will focus on trivial insulating states so that the gauged systems are exactly solvable, zero-correlation models exposing the underlying fractonic order. In fact all of the gauged models in these examples are stabilizer code Hamiltonians. Each of the examples has a bosonic analog in which the fermionic degrees of freedom correspond to spin-1/2 degrees of freedom and the subsystem fermionic parity symmetries correspond to subsystem Ising $Z_2$ symmetries. For instance, the fermionic 3D toric code is analogous in this sense to the ordinary 3D toric code.

\subsection{Global symmetry}

The general gauging procedure applied to global fermion parity symmetry on a square or cubic lattice reproduces the procedures discussed in Section \ref{sec:global}. The algebra of global fermion parity-preserving operators is generated by the on-site fermion parities $-i\gamma_i\gamma'_i$ and the parity hopping operators $\gamma_i\gamma_j$, hence the $Z_2$ gauge fields are attached to the lattice links. The transformations of Fig. \ref{fig:2Dgaugecv}a and Fig. \ref{fig:3Dgaugecv}a  correspond to particular choices of the antisymmetric function $T$. The 2D mapping corresponds to the function defined by
\begin{equation}
\begin{split}
    T(W,E)&=T(W,S)=T(S,E)=\\
    T(N,W)&=T(N,E)=T(N,S)=1
\end{split}
\end{equation}
and all other values not determined by antisymmetry 0. Here, $N$, $S$, $E$, and $W$ refer to the links directly north, south, east, and west of a given vertex, respectively. For the 3D mapping, these values are augmented with the additional definitions
\begin{equation}
\begin{split}
    T(O,I)&=T(O,S)=T(O,E)=\\
    T(N,O)&=T(W,O)=T(N,I)=\\
    T(S,I)&=T(E,I)=T(W,I)=1
\end{split}
\end{equation}
where $I$ and $O$ refer to the links into and out of the page relative to a given vertex.

\subsection{Linear symmetry}

As a simple example of subsystem symmetry, let us consider a system with one fermionic mode per site of a 2D square lattice in which the fermion parity along all $x$ and $y$ oriented lines is conserved. In this case the minimal couplings are four-body plaquette operators $\gamma_i\gamma_j\gamma_k\gamma_l$ where $ijkl$ is an elementary plaquette. To gauge the symmetries, we accordingly add a gauge qubit to each plaquette $p$ in the lattice, and enforce the Gauss's law constraint $-i\gamma_i\gamma'_i\prod_{p\ni i} X_p=1$ at every site. There are no local flux constraints that preserve this Gauss's law. However, the non-local constraints $O_L=\prod_{p\in L}Z_p$
do preserve the Gauss's law and thus are interpreted as linear subsystem symmetries of the gauged Hamiltonian. Here $L$ refers to any elementary row or column of plaquettes. This is similar to the case of global symmetries in 1D and linear subsystem symmetries in 2D, in which gauging the symmetry likewise yields subsystem symmetric states.\cite{GaugingSubsystem} In this case linear fermion parity symmetries are transmuted into linear $Z_2$ symmetries in a spin system.

If we begin with the trivial Hamiltonian $H_0$ of Eq. \ref{eq:H}, the gauged Hamiltonian $H_g=H_0$. The transformation of Eq. \ref{eq:transform} takes the form
\begin{equation*}
\begin{split}
    X_p&\to X_p\qquad Z_{(x,y)}\to c_{(x,y)} Z_{(x,y)} X_{(x-1,y+1)} X_{(x+1,y+1)} \\
    \gamma_i&\to \gamma_i\prod_{p\ni i} X_p\qquad\qquad \gamma_i'\to\gamma_i'
\end{split}
\end{equation*}
for a particular choice of the function $T$. Here, the 2D coordinates represent vertices of the \textit{dual} lattice, which correspond to plaquettes of the original lattice. As a result the linear symmetries are mapped to the operators
\begin{equation*}
\begin{split}
    O_x&=\prod_y X_{(x-1,y)}Z_{(x,y)}X_{(x+1,y)}\\
    O_y&=\prod_x Z_{(x,y)}
\end{split}
\end{equation*}
and $H_g$ is mapped to the subsystem symmetric spin Hamiltonian
\begin{equation}
    H_\text{lin}=-\sum_i\prod_{p\ni i}X_p
\end{equation}
whose ground space spontaneously breaks the symmetry. This model is equivalent to the strong-coupling limit of the Xu-Moore model.\cite{XuMoore} It would be interesting to explore what type of transition in the fermionic system corresponds to the subsystem symmetry breaking transition in the gauged spin system.

\subsection{Planar symmetry}
\label{sec:examplesC}

\subsubsection{Fermionic 2-foliated lineon model}

\begin{figure}
    \centering
    \includegraphics[width=0.4\textwidth]{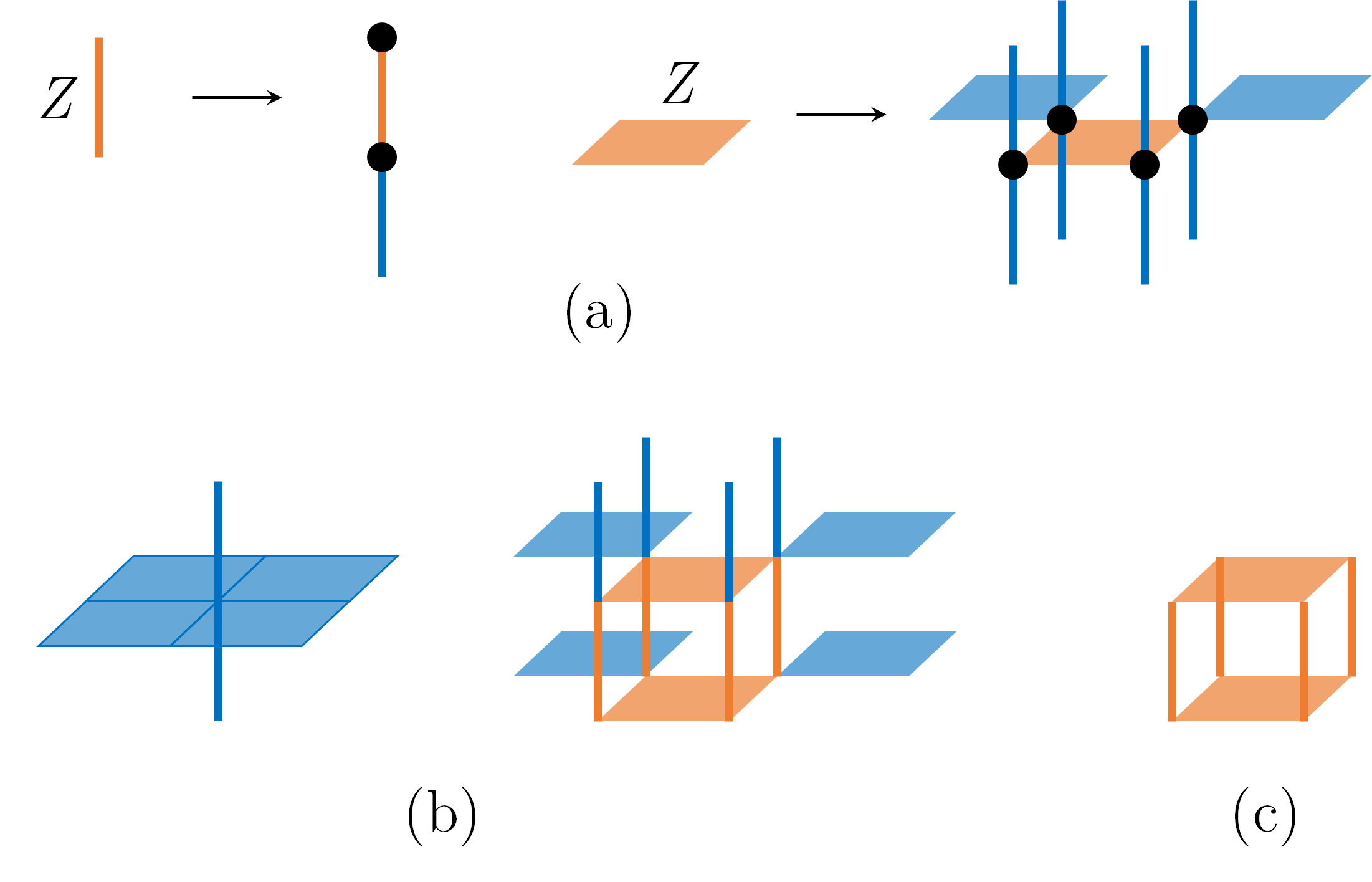}
    \caption{(a) Mapping of Pauli $Z$ operators under the transformation \ref{eq:transform} for a particular choice of antisymmetric function $T$, that takes $H_g$ of Eq. \ref{eq:Hg2F} to the fermionic 2-foliated lineon code Hamiltonian $H_\text{f2F}$, which is defined in a system with one qubit per $z$-oriented link and $xy$ plaquette. (b) Hamiltonian terms $A_v$ and $\tilde{B}_c$ of $H_\text{f2F}$. The terms are tensor products of Pauli operators. Blue denotes Pauli $X$ whereas orange denotes Pauli $Z$. (c) Cube term $B_c$ of $H_{2F}$ and $H_g$.}
    \label{fig:2folCv}
\end{figure}

Now we will consider systems that conserve planar fermion parity. First we will consider a trivial insulating Hamiltonian $H_0$ on a 3D cubic lattice in which fermion parity is conserved in $yz$ and $zx$ planes only. These symmetries act as $(-1)^{F_P}$ where $F_P$ is the number of fermions in plane $P$. The minimal couplings in this care are the two-body couplings $\gamma_i\gamma_{i+\hat{z}}$, and the plaquette operators $\gamma_i\gamma_j\gamma_k\gamma_l$ for plaquettes $ijkl$ normal to the $z$ direction. Following the general gauging procedure, $Z_2$ gauge fields are placed on every $z$-oriented link and every $z$-normal plaquette. The Gauss's law constraint is then imposed, which acts on the fermionic mode at site $i$ and the four plaquette gauge fields and two link gauge fields adjacent to $i$. The flux constraints $B_c$ added to the Hamiltonian are tensor products of Pauli $Z$ operators over the four $z$-oriented links and two $z$-normal plaquettes adjacent to cube $c$ (pictured in Fig. \ref{fig:2folCv}c). Thus the Hamiltonian after step 4 of the procedure has the form
\begin{equation}
    {H}_g=i\sum_i\gamma_i\gamma'_i -\sum_c B_c
    \label{eq:Hg2F}
\end{equation}

To complete step 5, we act with the local unitary circuit defined in Eq. \ref{eq:transform}, where the function $T$ is implicitly defined by the action of the circuit on Pauli $Z$ operators depicted in Fig. \ref{fig:2folCv}a. This transformation yields the stabilizer code spin Hamiltonian
\begin{equation}
    H_\text{f2F}=-\sum_v{A_v}-\sum_c\tilde{B}_c
\end{equation}
where $A_v$ is the emergent gauge symmetry associated with vertex $v$, and $\tilde{B}_c$ is the transformed flux constraint. Both of these terms are depicted in Fig. \ref{fig:2folCv}.

This model bears similarity to its bosonic analog, the 2-foliated lineon code introduced in Ref. \onlinecite{FractonStatistics}, whose Hamiltonian has the form
\begin{equation}
    H_\text{2F}=-\sum_v{A_v}-\sum_c{B}_c
\end{equation}
Both $H_\text{2F}$ and $H_\text{f2F}$ have two species of fractional lineon excitations, corresponding to emergent gauge charge and flux, which are excitations of the two types of Hamiltonian terms respectively. Whereas $H_\text{2F}$ has a self-duality that exchanges the two types of bosonic lineons, the lineonic gauge charges of $H_\text{f2F}$ correspond to the physical fermions of the original insulating state. Hence they exhibit fermionic exchange statistics, which will be demonstrated in Section \ref{sec:statistics}. In this sense $H_\text{f2F}$ is characterized by an emergent fermionic gauge theory of a fractonic nature. Nonetheless, it turns out that it is possible to transmute the statistics of lineonic gauge charge via attachment of lineonic flux. As a result, $H_\text{f2F}$ and $H_{2F}$ are actually related to one another by a finite depth circuit. This mapping will be discussed in detail in Section \ref{sec:unique}.

Note that the product of $\tilde{B}_c$ operators in a large rectangular prism is a `wireframe' operator which has support near the top and bottom faces and $z$ oriented edges of the rectangular prism. This operator represents a process of creation, movement, and annihilation of four fermionic lineon charges. Similarly, the product of $A_v$ operators in a large rectangular prism is a wireframe operator with the same geometry corresponding to the bosonic flux excitations.\cite{FractonStatistics}

\subsubsection{Fermionic X-cube model}
\label{sec:Xcube}

\begin{figure}
    \centering
    \includegraphics[width=0.42\textwidth]{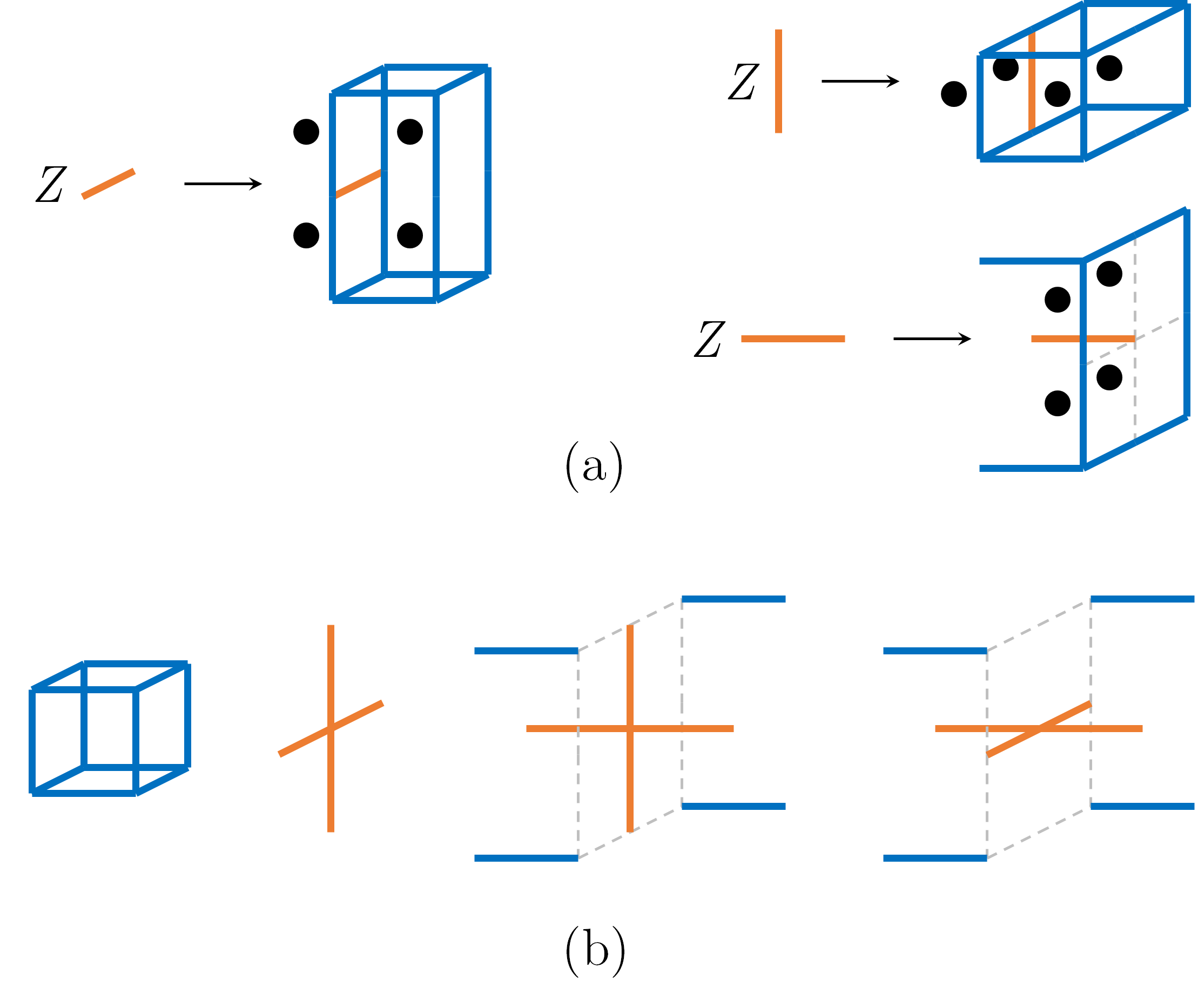}
    \caption{(a) Mapping of Pauli $Z$ operators under the transformation \ref{eq:transform} for a particular choice of antisymmetric function $T$ (augmented with an additional local unitary), that takes $H_g$ of Eq. \ref{eq:HgXC} to the fermionic X-cube Hamiltonian $H_\text{fXC}$. (b) Hamiltonian terms $A_c$, $\tilde{B}^{x}_v$, $\tilde{B}^{y}_v$, and $\tilde{B}^{z}_v$ of the fermionic X-cube model $H_\text{fXC}$, which is defined on a cubic lattice with one qubit per edge. The terms are tensor products of Pauli operators. Blue denotes Pauli $X$ whereas orange denotes Pauli $Z$.}
    \label{fig:fXCcv}
\end{figure}

Now we consider a trivial insulating system, with Hamiltonian $H_0$, on a cubic lattice with planar fermion parity symmetries in all three directions, $xy$, $yz$, and $zx$. This model is analogous to a spin-1/2 paramagnet with planar Ising symmetries, which yields the well-known X-cube model upon gauging\cite{Sagar16} (see Appendix \ref{app:Xcube}). Gauging the three sets of planar fermion parity symmetries of $H_0$ yields a variant of the X-cube model we call the fermionic X-cube model.

The minimal couplings in this case are the plaquette operators $\gamma_i\gamma_j\gamma_k\gamma_l$ where $ijkl$ is an elementary plaquette of the cubic lattice. Following the general procedure, a gauge qubit is placed on each plaquette $p$, and the Gauss's law constraint $-i\gamma_i\gamma'_i\prod_{p\ni i} X_p=1$ is imposed at every site. This Gauss's law acts on 12 gauge qubits adjacent to the site $i$. Four-body flux constraints $B_c^x$, $B_c^y$, and $B_c^z$ are then added to the Hamiltonian. We define $B_c^x=Z_pZ_qZ_rZ_s$ where $p$, $q$, $r$, and $s$ are the $xy$ and $zx$ faces of cube $c$, and likewise for $B_c^y$ and $B_c^z$. Thus, prior to the transformation of step 5, the gauged Hamiltonian has the form
\begin{equation}
    \label{eq:HgXC}
    {H}_g=i\sum_i\gamma_i\gamma'_i -\sum_c \left(B_c^x+B_c^y+B_c^z\right).
\end{equation}
It is helpful at this point to switch from the direct lattice to the dual lattice, in which the gauge qubits lie on the links as opposed to the plaquettes, the fermions reside at the center of elementary cubes, and the flux constraints $B_v^x$, $B_v^y$, and $B_v^z$ are now associated with vertices. We then act with the local unitary of Eq. \ref{eq:transform}, where the function $T$ is implicitly defined by the action on Pauli $Z$ operators defined in Fig. \ref{fig:fXCcv}. The resulting Hamiltonian, acting on the gauge qubits only, takes the form
\begin{equation}
    H_\text{fXC}=-\sum_cA_c -\sum_v \left(\tilde{B}_v^x+\tilde{B}_v^y+\tilde{B}_v^z\right)
\end{equation}
where $A_c$ is the emergent gauge symmetry term associated with cube $c$, and $\tilde{B}^\mu_v$ are the transformed flux constraints, both defined in Fig. \ref{fig:fXCcv}b. This model is an exactly solvable stabilizer code model bearing many similarities to the ordinary X-cube model, whose Hamiltonian for comparison is
\begin{equation}
    H_\text{XC}=-\sum_cA_c -\sum_v \left({B}_v^x+{B}_v^y+{B}_v^z\right).
\end{equation}
Like the X-cube model, the fractional excitations of the fermionic X-cube model consists of fracton gauge charges, which are excitations of the cube terms, and lineonic gauge flux, which are excitations of the vertex terms. Both fracton dipoles and lineon dipoles are planons free to move in the plane normal to the dipole axis. The braiding statistics of $H_\text{fXC}$ and $H_\text{XC}$ have the same structure of generalized Aharanov-Bohm phases between fractons and lineon dipoles, and lineons and fracton dipoles.

Although it is not possible to exchange pairs of fractons via a local process since individual fractons are immobile, it is possible to condense a stack of fracton dipoles, say with $z$-oriented dipolar axis, so that all of the fractons along a single $z$ axis collapse into a single lineonic superselection sector. If this condensation transition is induced in the fermionic X-cube model, the resulting lineons have fermionic exchange statistics, which suggests that the original fractons are themselves fermions. In fact, these lineons represent the lineonic gauge charge of a model in the same gapped phase as the fermionic 2-foliated lineon model $H_\text{f2F}$ introduced in the previous section (see Appendix \ref{app:condensation}). An important open question is whether there exists a statistical process that gives invariant meaning to the notion of fracton statistics in a given model, without appealing to the fate of the fracton under a phase transition. This question forms the basis of future work.

Despite the fermionic nature of the fracton excitations of $H_\text{fXC}$, it turns out that the fermionic X-cube model and X-cube model actually exhibit the same fractonic order, as discussed in Section \ref{sec:unique}. Indeed, it is possible to transmute the statistics of fractons by attaching gauge flux, allowing the construction of a finite depth quantum circuit which connects the two models.

\subsubsection{Half-fermionic X-cube model}
\label{sec:half}

Here we will describe a model that contains both fermionic and spin degrees of freedom with composite $Z_2$ planar subsystem symmetries. The gauged model is analogous to the ordinary X-cube model and the fermionic X-cube model discussed in the previous section, and will be called the half-fermionic X-cube model. The ungauged system is defined on a cubic lattice which is bipartitioned into two interpenetrating $A$ (even) and $B$ (odd) checkerboard sublattices. There is one fermionic degree of freedom on each $A$ sublattice site and one bosonic spin-1/2 degree of freedom on each $B$ sublattice site, and the planar symmetries have the form
\begin{equation}
    O_P=\prod_{i\in P\cap A}-i\gamma_i\gamma'_i\prod_{j\in P\cap B}X_j
\end{equation}
In other words $O_P$ measures the fermion parity in addition to flipping all spins in plane $P$. Note that the global fermion parity is generated by the subsystem symmetries, since the operators $O_\text{even}=\prod_{P\text{ even}}O_P$ and $O_\text{odd}=\prod_{P\text{ odd}}O_P$ correspond to global fermion parity and global Ising symmetries respectively. Therefore we expect to obtain a pure spin Hamiltonian upon gauging the symmetry.

As in the fermionic X-cube example, the minimal couplings for this model can be chosen to be four-body plaquette operators. On one sublattice they take the form $\gamma_i X_j\gamma_k X_l$ around plaquette $ijkl$ and on the other the form $X_{i'}\gamma_{j'}X_{k'}\gamma_{l'}$. These couplings generate the subsystem symmetric operator algebra together with the on-site symmetry representations $-i\gamma_i\gamma'_i$ and $X_i$. Hence, to gauge the symmetry a $Z_2$ gauge qubit is added to each plaquette, and the following Gauss's laws imposed:
\begin{equation}
    \begin{split}
    -i\gamma_i\gamma'_i\prod_{p\ni i} X_p=1\qquad i\in A\\
    X_i\prod_{p\ni i} X_p=1\qquad i\in B
    \end{split}
\end{equation}
The terms $B_c^x$, $B_c^y$, and $B_c^z$, defined the same as in Eq. \ref{eq:HgXC}, are then added to the Hamiltonian to gap out flux excitations. Thus the gauged Hamiltonian on the constrained Hilbert space takes the form
\begin{equation}
    {H}_g=i\sum_{i\in A}\gamma_i\gamma'_i-\sum_{i\in B}X_i -\sum_c \left(B_c^x+B_c^y+B_c^z\right)
\end{equation}

Description of the local unitary transformation to a tensor product Hilbert space and the resulting spin Hamiltonian $H_\text{hXC}$, dubbed the half-fermionic X-cube model, is relegated to Appendix \ref{app:halfXcube} due to its complexity. $H_\text{hXC}$, has the same structure of fractonic gauge charge and lineonic gauge flux excitations as $H_\text{XC}$ and $H_\text{fXC}$. However, whereas both of these models (as well as $H_\text{2F}$ and $H_\text{f2F}$) exhibit planon gauge charges that are exclusively bosonic, this model harbors fermionic planon gauge charge (see Section \ref{sec:statistics}). It is characterized by an emergent fractonic gauge theory of a mixed fermionic and bosonic nature.


\subsection{Fractal subsystem symmetry: Fermionic Fibonacci prism model}

\begin{figure}
    \centering
    \includegraphics[width=0.4\textwidth]{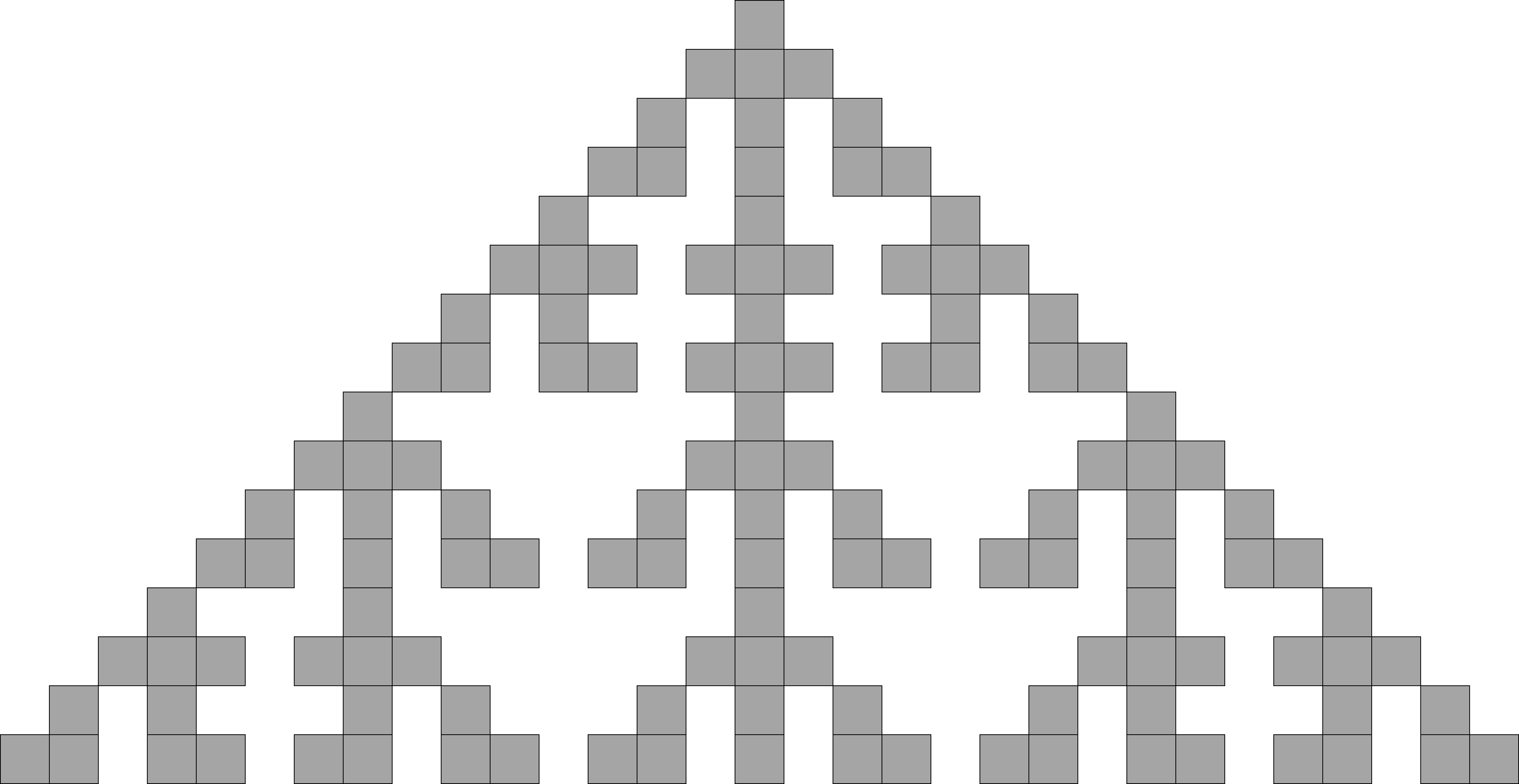}
    \caption{A finite Fibonacci triangle fractal.}
    \label{fig:Fibonacci}
\end{figure}

\begin{figure}
    \centering
    \includegraphics[width=0.45\textwidth]{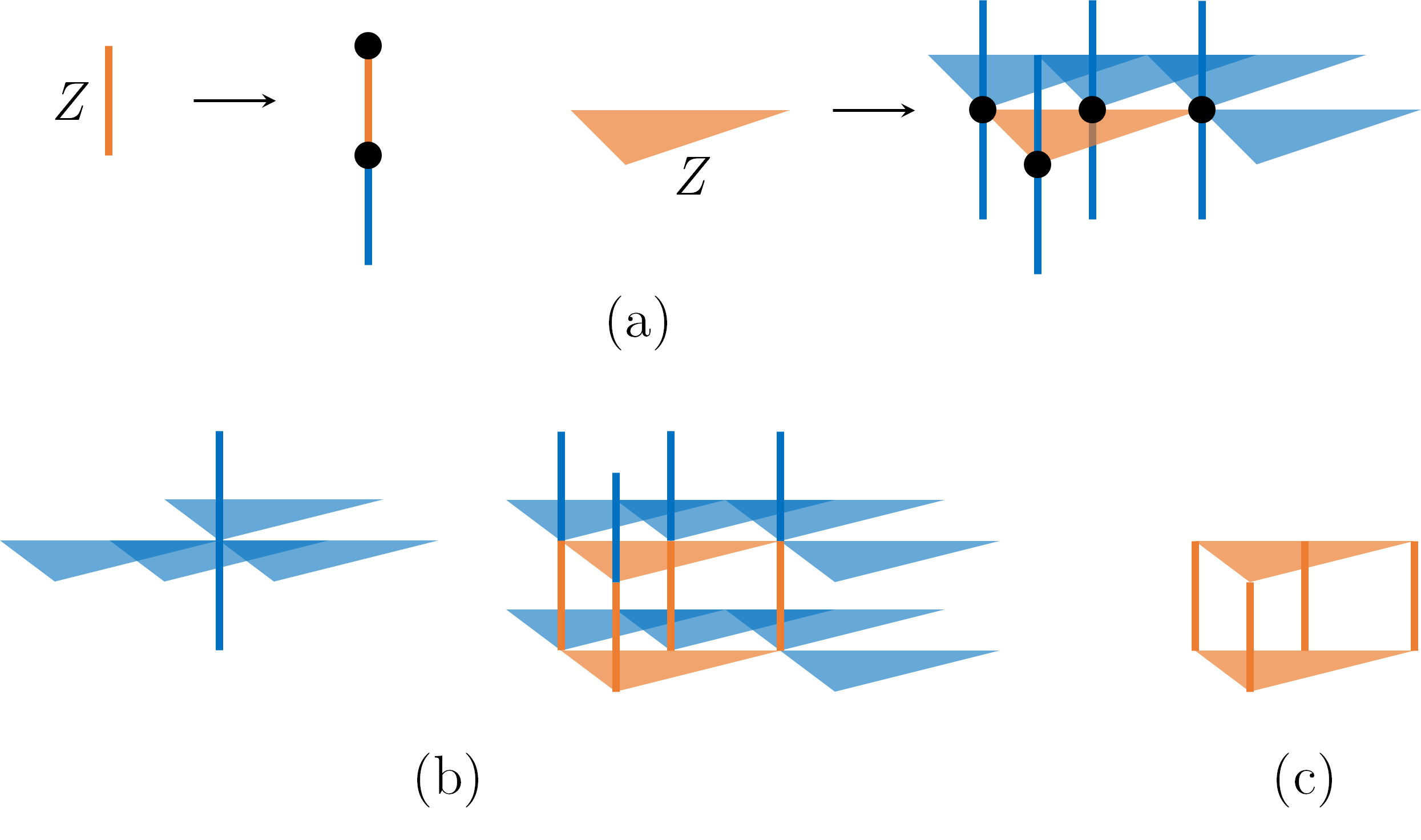}
    \caption{(a) Mapping of Pauli $Z$ operators under the transformation \ref{eq:transform} for a particular choice of antisymmetric function $T$, that takes $H_g$ of Eq. \ref{eq:HgFib} to the fermionic Fibonacci prism model $H_\text{fFib}$. (b) The vertex and prism terms $A_v$ and $\tilde{B}_p$ of $H_\text{fFib}$, which are tensor products of Pauli operators. Blue denotes Pauli $X$ whereas orange denotes Pauli $Z$. (c) Prism term $B_p$ of $H_{Fib}$ and $H_g$.}
    \label{fig:Fib}
\end{figure}

In this section we illustrate how the gauging procedure works for fractal subsystem fermion parity symmetries via a simple example. In particular, we consider a system with one fermionic mode per site of a 3D cubic lattice, in which fermion parity is conserved in each subsystem with the geometry of a stack of `Fibonacci triangle' fractals of infinite extent. The Fibonacci triangle fractal can be defined as the spacetime support of a 1D cellular automaton;\cite{YoshidaFractal} a finite version is pictured in Fig. \ref{fig:Fibonacci}. The symmetry group is most easily defined as the group of subsystem fermion parity symmetries of the following Hamiltonian:
\begin{equation}
\begin{split}
    H^\text{Fib}_0=H_0+J_1\sum_{i}i\gamma_i\gamma_{i+\hat{z}}\\+J_2\sum_{i}\gamma_i\gamma_{i-\hat{x}+\hat{y}}\gamma_{i+\hat{y}}\gamma_{i+\hat{x}+\hat{y}}
\end{split}
\end{equation}
Since the $J_1$ term hops fermion parity along the $z$ direction, all subsystem fermion parity symmetries must act on entire chains of fermionic modes parallel to the $z$ axis. Conversely the $J_2$ term ensures that all symmetries must act on the entirety of infinite Fibonacci triangle planar subsystems oriented along the $xy$ plane. Therefore, the symmetries of $H^\text{Fib}_0$ are stacked Fibonacci triangle subsystem fermion parity symmetries, and the minimal couplings are by construction the $J_1$ and $J_2$ terms of the Hamiltonian. A careful analysis of subsystem symmetries with this geometry in spin systems was conducted in Ref. \onlinecite{DevakulFractal}.

Following the general gauging procedure, we gauge the Hamiltonian $H^\text{Fib}_0$ by introducing gauge qubits attached to each $z$-oriented lattice link and to each isosceles triangle enclosed by the support of a $J_2$ four-body coupling. We then enforce the generalized Gauss's law, which acts on the two link qubits and four isosceles triangle qubits adjacent to a given vertex (in addition to the fermionic mode at that vertex). The flux constraints $B_p$ are six-body terms associated with each isosceles triangular prism $p$, and take the form depicted in Fig. \ref{fig:Fib}c, yielding the Hamiltonian
\begin{equation}
    \label{eq:HgFib}
    {H}_g=i\sum_{i\in A}\gamma_i\gamma'_i-\sum_pB_p
\end{equation}
in the $J_1,J_2\to0$ limit. After the local unitary transformation of step 5, whose action is depicted in Fig. \ref{fig:Fib}a for a particular choice of $T$, the gauged Hamiltonian has the form
\begin{equation}
    H_\text{fFib}=-\sum_vA_v-\sum_p\tilde{B}_p
\end{equation}
where the emergent gauge symmetry $A_v$ and the transformed flux constraint $\tilde{B}_p$ are depicted in Fig. \ref{fig:Fib}b. $H_\text{fFib}$ bears similarity to its bosonic analog
\begin{equation}
    H_\text{Fib}=-\sum_vA_v-\sum_p{B}_p
\end{equation}
which is called the Fibonacci prism model and belongs to the class of fractal spin liquids introduced by Yoshida in Ref. \onlinecite{YoshidaFractal}. Both $H_\text{Fib}$ and the fermionic Fibonacci prism model $H_\text{fFib}$ have both lineonic charge and flux excitations, which correspond to violations of the two types of stabilizer generators. The fusion rules of the two sectors are spatial inverses of one another in both model. However, the charge excitations of $H_\text{Fib}$ are bosonic, giving rise to a self-duality in the model that is not present in $H_\text{fFib}$, which has fermionic lineon charge excitations as we will see in the following section. As we will argue in Section \ref{sec:unique}, the fractonic orders exhibited by $H_\text{fFib}$ and $H_\text{Fib}$ are unique from one another. Thus $H_\text{fFib}$ represents a genuinely novel fractonic order characterized by an emergent fermionic gauge theory.

\section{Emergent fermions and gauging correspondence}\label{sec:statistics}

In this section we will discuss the general correspondence between subsystem fermion parity symmetry and the fractonic order obtained upon gauging the symmetry. Under this correspondence the mobility and statistics of the gauge charge excitations are fully determined by the way the symmetry charges transform under the subsystem symmetries and the global fermion parity. As an example, let us review the correspondence in the case of the ordinary X-cube model.\cite{GaugingSubsystem}

\subsection{Review: X-cube model}
\begin{figure}
    \centering
    \includegraphics[width=0.47\textwidth]{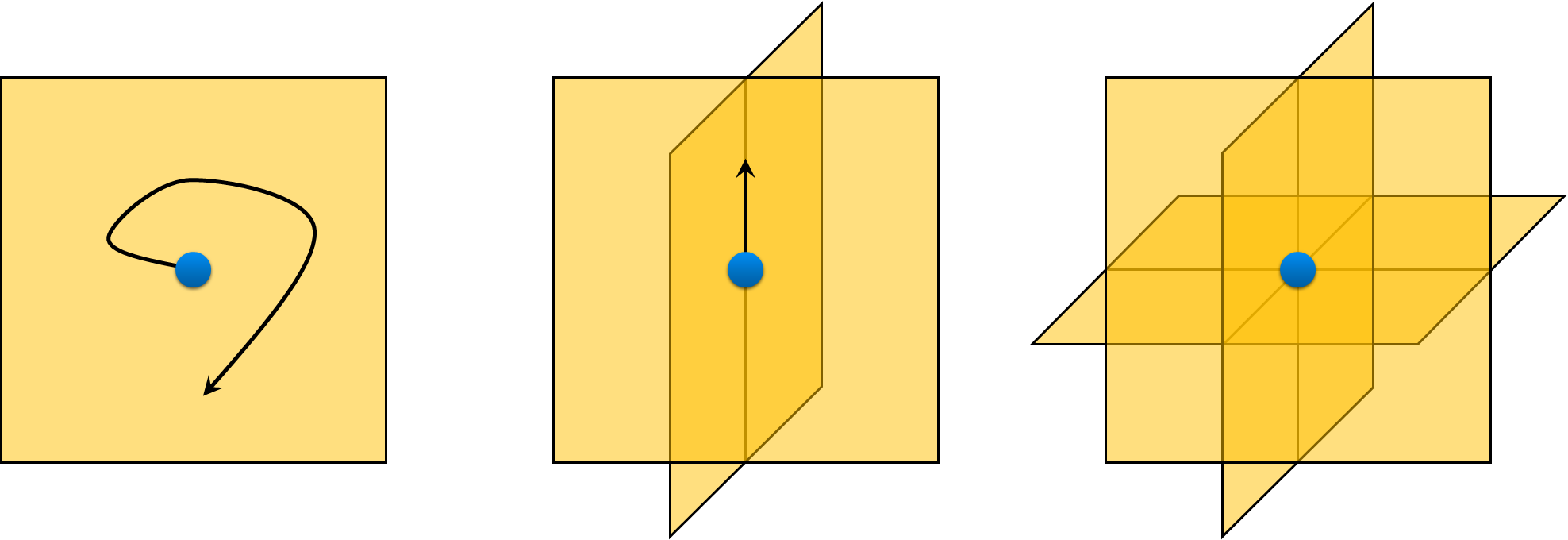}
    \caption{Symmetry charges transforming under planar symmetries in one, two, or three directions are planons, lineons, and fractons respectively. Gauge charges in the corresponding fractonic order inherit these mobility constraints.}
    \label{fig:charges}
\end{figure}

As detailed in Appendix \ref{app:Xcube}, the X-cube model is obtained by gauging three directions of planar Ising symmetries in a paramagnetic system with one spin-1/2 degree of freedom on each site of a cubic lattice. Each symmetry charge in this paramagnet transforms under three intersecting, mutually perpendicular planar symmetries. Therefore, as long as the symmetry is preserved an individual symmetry charge is pinned at the intersection of these three planes. Upon gauging, these charges become the fractons in the X-cube model. Conversely, symmetry charge dipoles, i.e. pairs of adjacent symmetry charges, transform only under (two) planar symmetries normal to their dipole moment. Therefore, they become planons upon gauging; in particular they map onto the fracton dipoles of the X-cube model. If only two directions of planar symmetries are preserved, say $yz$ and $zx$, then individual symmetry charges can be moved in the $z$ direction only, hence they become lineons upon gauging the symmetry. In this 2-foliated system, symmetry charge dipoles with $x$ or $y$ oriented dipole moment are planons via the same mechanism as the 3-foliated system. The fracton model obtained by gauging is the 2-foliated lineon model introduced in Ref. \onlinecite{FractonStatistics}. This correspondence between symmetry charge and fractonic gauge charge is summarized in Fig. \ref{fig:charges}.

\subsection{Fermionic X-cube and 2-foliated models}

The correspondence is similar in the case of planar fermion parity symmetry. In the three models introduced in Section \ref{sec:examplesC}, the structure of gauge charge mobility is identical to the respective bosonic analogs discussed above, owing to the fact that the symmetry groups are structurally identical. In particular, in the fermionic and half-fermionic X-cube models, the elementary symmetry charges become fractons upon gauging, whereas in the fermionic 2-foliated lineon model, they become lineons. However, because the symmetry charges in the ungauged fermionic models are fermions, i.e. they transform under global fermion parity, they correspond to emergent fermionic excitations after gauging. (In the half-fermionic model, some are fermions and some are bosons as the name suggests).
On the other hand, the symmetry charge dipoles in the ungauged fermionic X-cube model and fermionic 2-foliated models are bosons since they do not transform under global fermion parity, whereas in the ungauged half-fermionic X-cube model they are composed of one bosonic charge and one fermionic charge and are thus fermions. These dipoles become bosonic and fermionic planon gauge charges respectively upon gauging the symmetry.

The fermionic statistics of symmetry charges manifest in the exchange statistics of the mobile gauge charges after gauging, which are robust universal properties of the resulting fractonic order. It is well-known that exchange statistics of two-dimensional particles, i.e. anyons or planons, are well-defined in gapped systems and can be computed by exchanging two identical particles in a way such that the non-universal geometric phases arising in the process fully cancel.\cite{LevinWen,KitaevAnyons} Similar processes have been introduced that demonstrate the universal nature of lineon exchange statistics.\cite{FractonFusion,SongTwistedFracton} In the following, we will demonstrate the gauging correspondence by computing the exchange statistics of the planon and lineon gauge charges in the fermionic and half-fermionic X-cube models, and the fermionic 2-foliated model.

On the other hand, the notion of fracton statistics is quite subtle because fractons are fundamentally immobile as individual excitations, so it is impossible to exchange a pair of fractons via a local process. As discussed in Section \ref{sec:Xcube}, the fermionic nature of fracton excitations of $H_\text{fXC}$ is revealed by driving a phase transition in which one stack of fracton dipoles is condensed, consequently transforming individual fractons into fermionic lineons of a model in the same phase as $H_\text{f2F}$. It is natural to expect the fermionicity of fractonic gauge charge in the fermionic and half-fermionic X-cube models to manifest in a more direct way. A natural question to ask is whether there exist local statistical processes that generalize the notion of identical particle exchange to fractonic fractional excitations. This question forms the basis of future work; in the present we consider lineon and planon exchange statistics.

\subsubsection{Planon and lineon exchange statistics}

\begin{figure}
    \centering
    \includegraphics[width=0.48\textwidth]{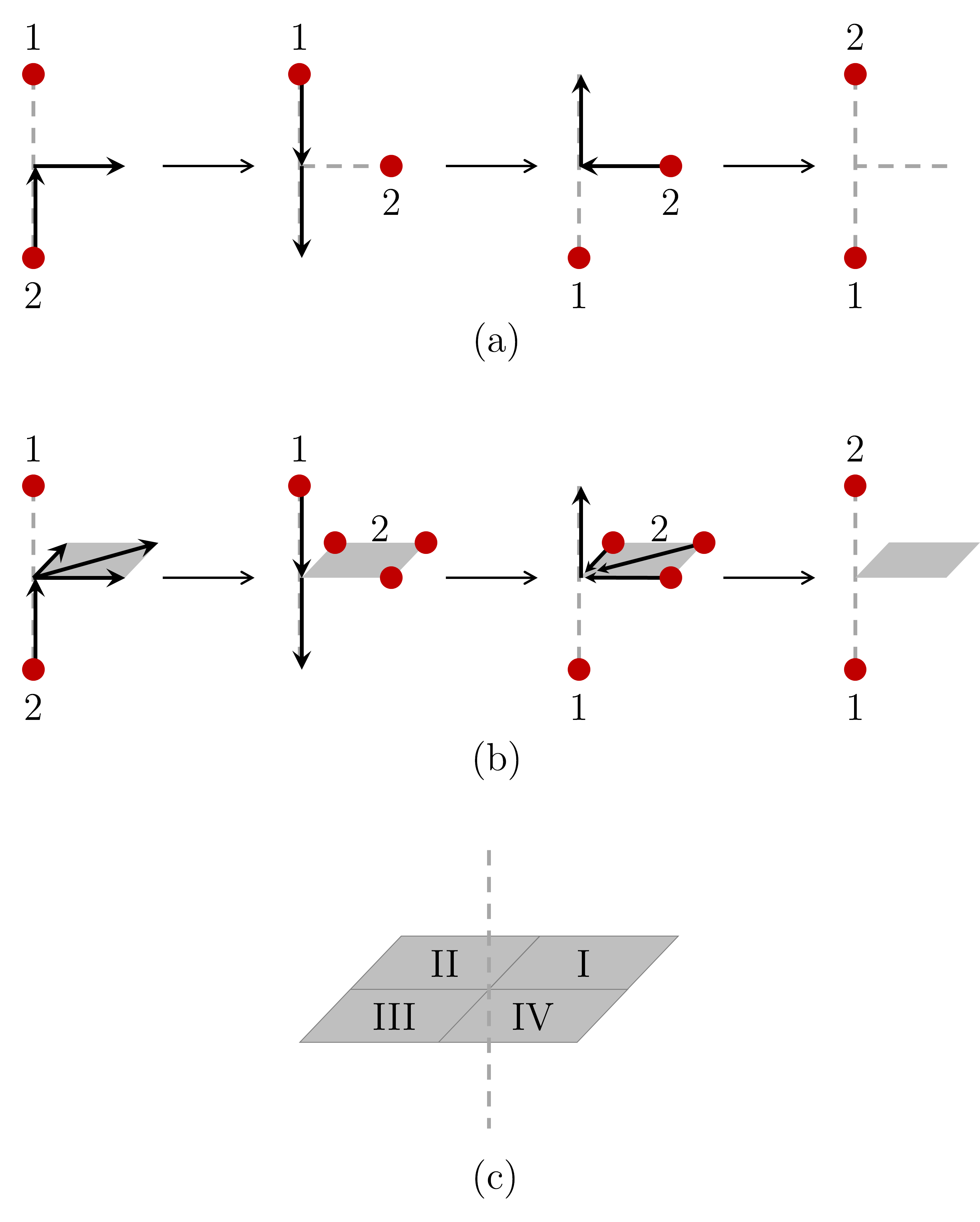}
    \caption{(a) A process that exchanges a pair of identical anyons (in 2D) or planons (in 3D) such that all non-universal phase factors cancel, revealing the invariant nature of the exchange statistic. (b) A process that exchanges a pair of identical lineons in a 2-foliated model. In the first step, lineon 2 is moved halfway, and then out of the way via a membrane operator which transforms the excitation into three new lineon excitations at the other corners. Note that, although the energy of this intermediate state is higher, the bound state of three lineons lies in the same superselection sector as lineon 1 and 2. In the next step, lineon 1 is moved all the way down, and finally the triple of lineons is transformed back into lineon 2 and moved the rest of the way up, completing the process. (c) Four inequivalent processes of lineon exchange, corresponding to the four quadrants that lineon 2 may be moved into.}
    \label{fig:exchange}
\end{figure}

\begin{figure}
    \centering
    \includegraphics[width=0.35\textwidth]{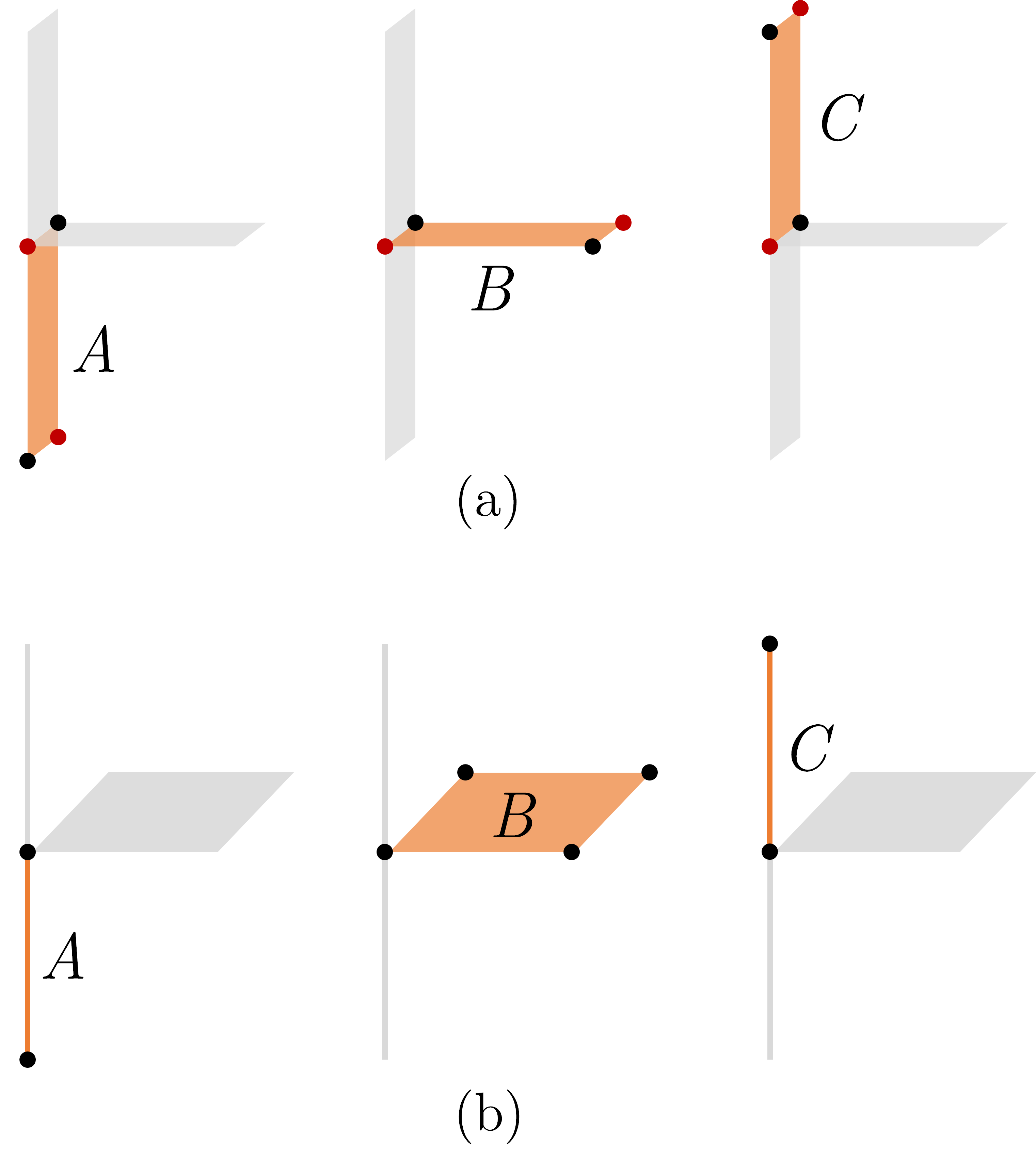}
    \caption{Operators $A$, $B$, and $C$ that realize the exchange process $O=CB^\dagger A^\dagger C^\dagger BA$ of (a) fracton dipoles in the fermionic or half-fermionic X-cube models and (b) lineons in the fermionic 2-foliated lineon code. These operators act on the Gauss's law-constrained Hilbert spaces prior to the transformation of step 5 in Section \ref{sec:subsystem}. In (a), we have reverted to the direct lattice in which gauge qubits lie on plaquettes. Here, orange represents Pauli $Z$ on the gauge qubits, whereas the black dots represent the Majorana operator $\gamma$ at a particular site. The red dots also represent $\gamma$ for the fermionic X-cube model, but for the half-fermionic X-cube model they represent Pauli $Z$ acting on matter spins living on the $B$ sublattice. Likewise in (b), orange represents Pauli $Z$ on gauge qubits and the black dots represent $\gamma$.}
    \label{fig:exchange2}
\end{figure}

The essential principle underlying nontrivial emergent exchange statistics is the encoding of statistics in the commutation relations of hopping operators that create particle-antiparticle pairs in adjacent locations.\cite{LevinWen,KitaevAnyons,FractonFusion} For emergent fermions, these operators anticommute if acting simultaneously on a particular site. For $z$-mobile lineons in 2-foliated models, there are rectangular membrane operators parallel to the $xy$ plane that create sets of four lineons at its corners, in addition to hopping operators in the $z$ direction. These generalized hopping operators are generically composed of elementary local steps which are defined up to an arbitrary phase. However, it is possible to exchange two identical particles such that these phases cancel, as depicted in Fig. \ref{fig:exchange}a for both planons and lineons. These processes can be understood heuristically as composed of three steps: 1) the first particle is moved halfway and then out of the way, 2) the second particle is moved all the way, and finally 3) the first particle is moved back to the original path and then the rest of the way. In each case the total exchange process corresponds to the action of an operator
\begin{equation}
    O=CB^\dagger A^\dagger C^\dagger BA
\end{equation}
where $A$, $B$, and $C$ are particle creation operators in a given model. In general, $O$ acts as the identity on a starting state with an excitation at each of the two locations, up to the universal exchange phase $\theta$. Figure \ref{fig:exchange2} depicts movement operators for fracton dipole exchange in the fermionic and half-fermionic X-cube models, and lineon charge exchange in the fermionic 2-foliated model. To write these operators, it is most natural to revert to the original Hilbert space containing both $Z_2$ gauge fields and fermionic matter degrees of freedom, subject to the generalized Gauss's law constraints equating the fermion number with the gauge field configuration at each site. For the operators defined in Fig. \ref{fig:exchange}b on this Hilbert space, we find that $\{A,B\}=\{B,C\}=\{C,A\}=0$ for planons in the half-fermionic X-cube model and lineons in the fermionic 2-foliated model, due to the Majorana fermion anticommutation relations---hence $O=-1$ meaning these excitations are fermions. Note that for lineons, the result is independent of which of the four inequivalent exchange processes depicted in Fig. \ref{fig:exchange} is used. On the other hand, $[A,B]=[B,C]=[C,A]=0$ for planons in the fermionic X-cube and 2-foliated models, so $O=1$ implying these quasiparticles are bosons. These results are what we expected according to the gauging correspondence.

\subsection{Fermionic Fibonacci prism model}

\begin{figure}
    \centering
    \includegraphics[width=0.48\textwidth]{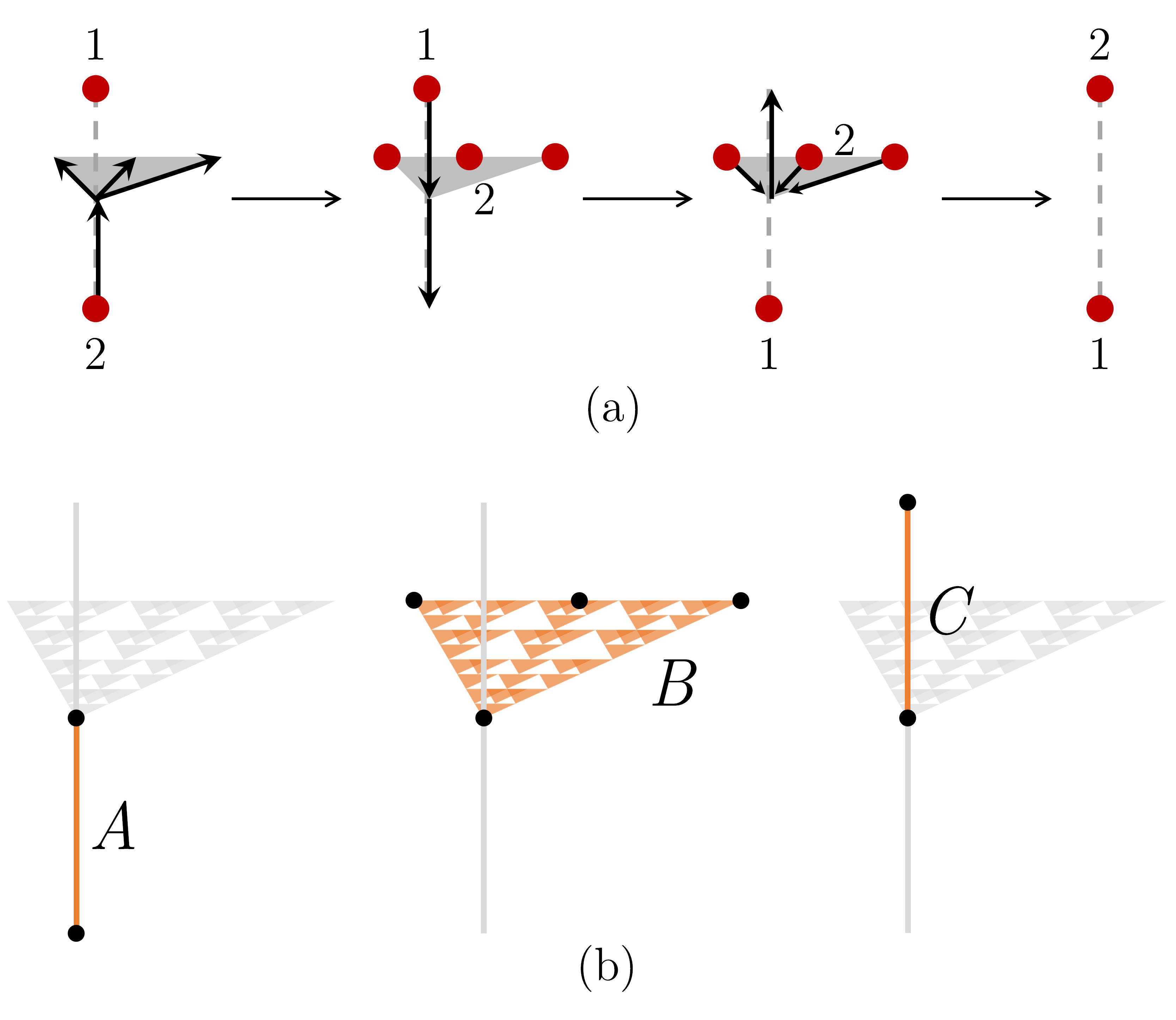}
    \caption{(a) A process that exchanges a pair of identical lineons in Fibonacci prism models such that all non-universal phase factors cancel, revealing the invariant nature of the exchange statistic. (b) Operators $A$, $B$, and $C$ that realize this exchange process as the total operator $O=CB^\dagger A^\dagger C^\dagger BA$. These operators act on the Gauss's law-constrained Hilbert spaces prior to the transformation of Eq. \ref{eq:transform} in Section \ref{sec:subsystem}. Orange represents the action of Pauli $Z$ on gauge qubits whereas the black dots represent the Majorana operator $\gamma$ at a particular site. Note that $B$ is a membrane operator whose support has the geometry of a discrete Fibonacci triangle fractal.}
    \label{fig:Fibexchange}
\end{figure}

The gauging correspondence works in a similar manner in the case of fractal subsystem symmetries. For the fermionic Fibonacci prism model, the starting point is a model that conserves fermion parity in every subsystem with the geometry of a stack of infinite Fibonacci triangle fractals (lying parallel to the $xy$ plane). Each individual symmetry charge transforms under many of these symmetries, such that it can only move in the $z$ direction while preserving all of these symmetries. The corresponding gauge charge is therefore a lineon. However, certain ($xy$) planar configurations of symmetry charge, in particular, sets of four symmetry charges at positions $i$, $i-L\hat{x}+L\hat{y}$, $i+L\hat{y}$, and $i+L\hat{x}+L\hat{y}$ with $L=2^n$, transform trivially under all symmetries. Therefore, lineon charges in the gauged model can be created in sets of four by a triangular membrane operator of this shape and size. These membrane operators actually have a fractal geometry: they act on a set of gauge fields forming a Fibonacci triangle.

The lineon symmetry charge is a fermion in the ungauged model. Therefore we expect the corresponding gauge charge to have fermionic exchange statistics. Indeed, there is a statistical exchange process very similar in spirit to the lineon and planon exchange processes discussed for planar fractonic orders in the previous section that gives this exchange statistic an invariant meaning. This process and the movement operators $A$, $B$, and $C$ that comprise it are depicted in Fig. \ref{fig:Fibexchange}. Note that since $B$ is a membrane operator with a fractal geometry, it creates intermediate excited states with energy up to $O(\log L)$ where $L$ is the length of the membrane.\cite{} However each of these excited states belongs to the same superselection sector as lineon 2. The overall process corresponds to the composite operator $O=CB^\dagger A^\dagger C^\dagger BA$. We find that $O=-1$ due to mutual anticommutation of the movement operators. Therefore, the lineons in the fermionic Fibonacci prism model are emergent fermions.

\section{Statistical transmutation and relations between fractonic orders}
\label{sec:unique}

In this section we address the question of whether the fractonic orders obtained by gauging subsystem fermion parity symmetries of trivial insulating states in Section \ref{sec:examples} are unique as fractonic orders from their analogs obtained by gauging $Z_2$ subsystem symmetries in Ising paramagnets. We regard two models as representative of the same fractonic order if and only if there is a local unitary transformation (i.e. a finite depth quantum circuit) that maps between the ground spaces of the two models.\cite{Xie10} Note that this is a much stricter equivalence relation than the foliated fracton equivalence.\cite{3manifolds,FractonEntanglement,MajoranaCheckerboard,Checkerboard} Under this definition, we find that the fermionic X-cube model has the same fractonic order as the ordinary X-cube model. Moreover, the fermionic 2-foliated lineon model has the same order as the bosonic 2-foliated lineon model. To this end, we construct local unitary Clifford circuits that map between these pairs of models. On the other hand, we argue that the fermionic Fibonacci fractal prism model is distinct as a fractonic order from its bosonic analog.

For comparison, gauging global fermion parity in 2D yields the same topological order as gauging $Z_2$ symmetry in a 2D Ising paramagnet---that of the 2D toric code, where the $\epsilon$ fermion is interpreted as either a gauge charge or a gauge flux, respectively. Conversely, gauging 3D global fermion parity yields a distinct topological order from gauging $Z_2$ symmetry in a 3D Ising paramagnet, distinguished by the respective statistics of their deconfined gauge charges. The difference between 2D and 3D arises because in 2D both $Z_2$ charge and flux are created by string operators, hence it is possible to transmute the charge statistics from bosonic to fermionic by attaching $\pi$-flux. Conversely, in 3D gauge flux excitations form closed loops, hence there is no way to attach flux to charge to transmute the statistics.\cite{LevinWen} In the planar fracton models constructed in Section \ref{sec:examplesC}, it turns out it is possible to transmute the statistics of all gauge charges to render them bosonic, by attaching the appropriate gauge fluxes. Let us illustrate this statistical transmutation in detail in the case of lineonic gauge charge in the fermionic 2-foliated model.

\subsection{Fermionic 2-foliated model}

\begin{figure}
    \centering
    \includegraphics[width=0.4\textwidth]{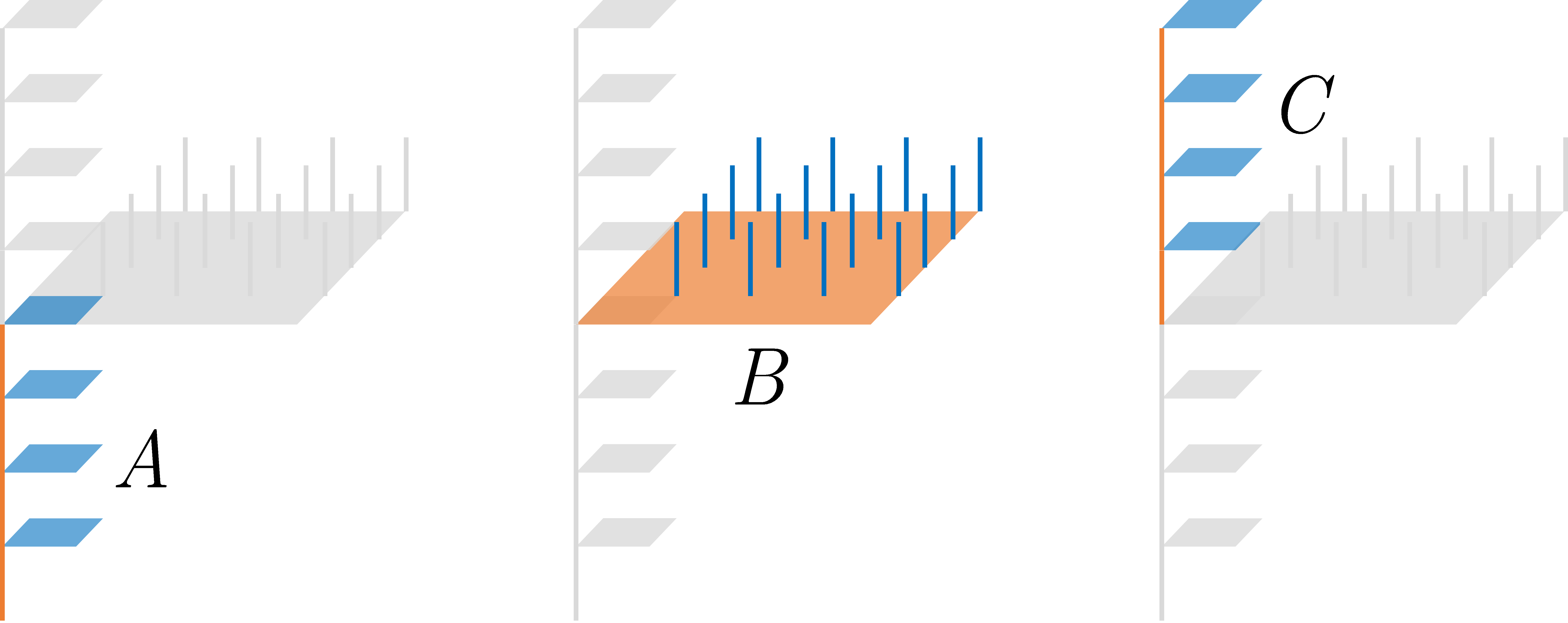}
    \caption{Operators $A$, $B$, and $C$ that realize the lineonic exchange process $O=CB^\dagger A^\dagger C^\dagger BA$ of bound states of lineonic charge and flux in the bosonic 2-foliated lineon code $H_\text{2F}$.}
    \label{fig:boundstate}
\end{figure}

\begin{figure}
    \centering
    \includegraphics[width=0.48\textwidth]{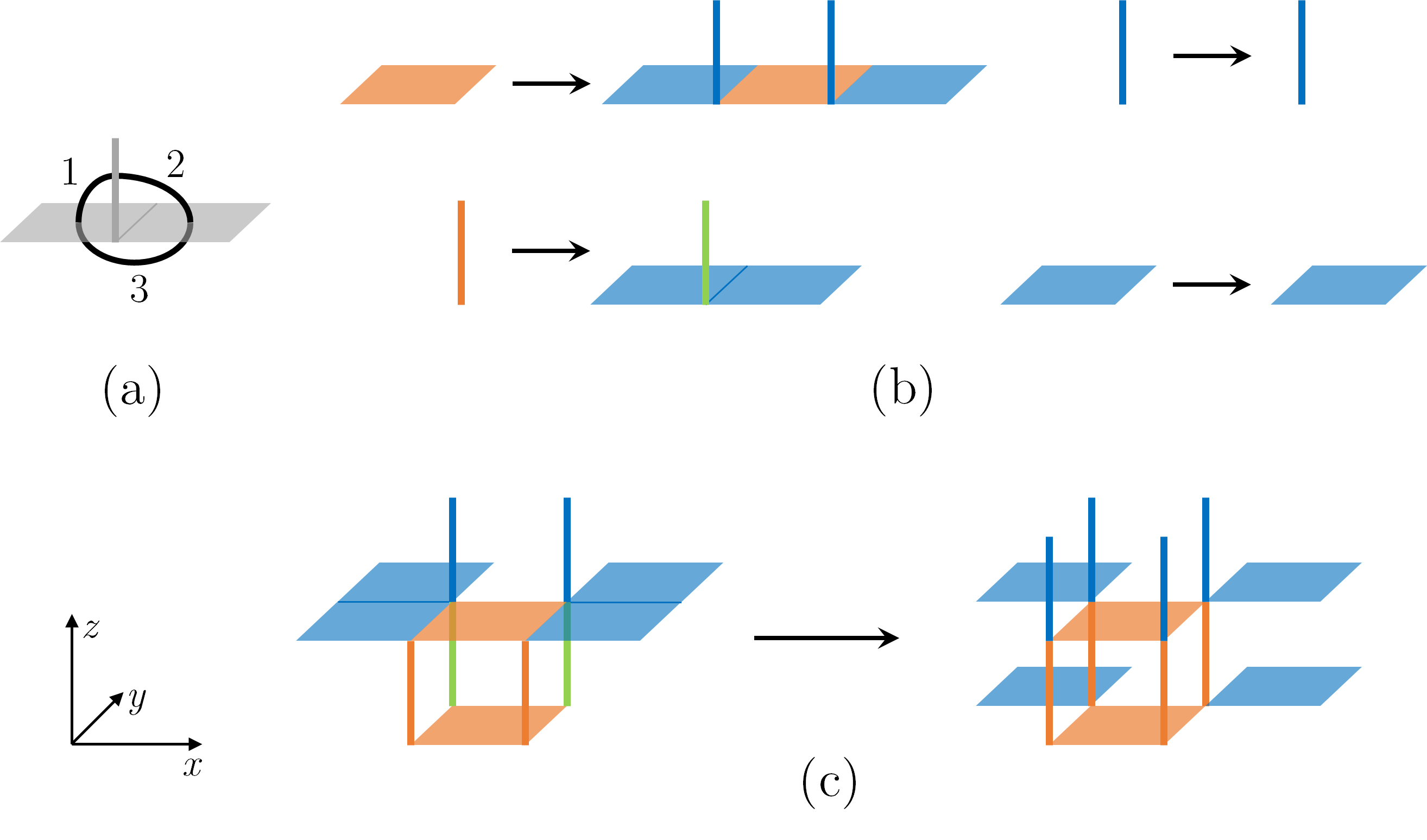}
    \caption{(a) Unit cell of a translation-invariant unitary circuit that maps the ground space of $H_\text{2F}$ to that of $H_\text{f2F}$. Depicted are three instances of the two-qubit gate $HCZH$, where $H$ is Hadamard rotation and $CZ$ is the controlled-$Z$ gate. (b) Action of the circuit on the operator algebra. Here blue, green, and orange represent Pauli $X$, $Y$, and $Z$ respectively. (c) The stabilizer generator $A_{(x,y,z)}A_{(x+1,y,z)}B_{(x+1/2,y-1/2,z-1/2)}$ of the stabilizer code defined by $H_\text{2F}$ is mapped to the flux constraint cube term $\tilde{B}_{(x+1/2,y-1/2,z-1/2)}$ of $H_\text{f2F}$.}
    \label{fig:f2folto}
\end{figure}

We will consider statistical transmutation from bosonic to fermionic lineon gauge charge. In the bosonic charge 2-foliated lineon model $H_\text{2F}$, the elementary gauge flux and charge are both lineons mobile in the $z$ direction. Charge lives on the vertices of the direct lattice whereas flux lives on the vertices of the dual lattice. In fact the model is self-dual under exchange of charge and flux.

The elementary flux attachment consists of binding a single lineonic flux to each lineonic gauge charge. Suppose we bind to each charge a flux lying on an axis positioned in the positive $x$ and $y$ direction relative to the charge. Let us compute the exchange statistics of this bound state using the process of Fig. \ref{fig:exchange}. Depending on which of the four inequivalent processes in Fig. \ref{fig:exchange}c (exchange in quadrant I, II, III, or IV) are used to exchange the composite lineons, either a $+1$ or $-1$ exchange statistic will be obtained. As an example the movement operators $A$, $B$, and $C$ are defined in Fig. \ref{fig:boundstate} for exchange in quadrant I. Processes I and III yield a $-1$ statistic since $\{AC,B\}=[A,C]=0$ in these cases whereas processes II and IV yield a $+1$ statistic since $[A,B]=[B,C]=[A,C]=0$ in these cases. The different in result is not because the exchange statistics are ill-defined; it is simply because unlike anyons or planons, lineons are characterized by four inequivalent lineon exchange statistics, each of which is a robust universal property of the underlying fractonic order. In the case of a lineonic flux-charge bound state, the four exchange statistics do not have a uniform value. In other words, this bound state is neither a fermion nor a boson; it represents a new kind of quasiparticle statistics that is unique to lineons. The lineon excitations of the semionic X-cube model\cite{MaLayers} are another example of this kind of statistics.

This calculation tells us that in order to transmute the lineon exchange statistics into pure fermionic statistics, we must attach more than one elementary lineonic flux. In fact, binding one additional lineonic flux in the negative $x$, positive $y$ direction will further transmute the statistics of processes II and IV to a $-1$ while leaving intact the statistics of processes I and III, which can be verified by examining the same exchange process for this particular bound state. Therefore, attaching to each lineonic gauge charge $(x,y)$ the dipolar bound state of lineonic fluxes $(x-1/2,y+1/2)$ and $(x+1/2,y+1/2)$, which is itself a planon, transmutes the statistics of each charge into a pure fermion. (Here the lineons are labelled by the $(x,y)$ coordinate of their axis of mobility). Moreover, it can be checked that this transmutation preserves all of the braiding statistics in the model, including the trivial braiding statistic between adjacent lineon dipole charges.

Therefore, by redefining gauge charge via attachment of flux, it is possible to rearrange the fusion and statistics structure of $H_\text{2F}$ so that it is isomorphic to the structure of the fermionic 2-foliated lineon model $H_\text{f2F}$. This suggests the existence of a local unitary circuit that maps between the ground spaces of the two models. Indeed, Fig. \ref{fig:f2folto} depicts such a circuit. Note that this circuit maps the stabilizer generators in a way that is consistent with the expectation from flux attachment. In particular, the stabilizer generator $A_{(x,y,z)}A_{(x+1,y,z)}B_{(x+1/2,y-1/2,z-1/2)}$ of $H_\text{2F}$ (Fig. \ref{fig:f2folto}c), whose corresponding wireframe operator represents movement of bound states of lineonic charge and planonic flux, is mapped to the stabilizer $\tilde{B}_c$ of $H_\text{f2F}$, whose wireframe operator represents movement of fermionic lineon charge.

\subsection{Fermionic and half-fermionic X-cube models}

\begin{figure}
    \centering
    \includegraphics[width=0.45\textwidth]{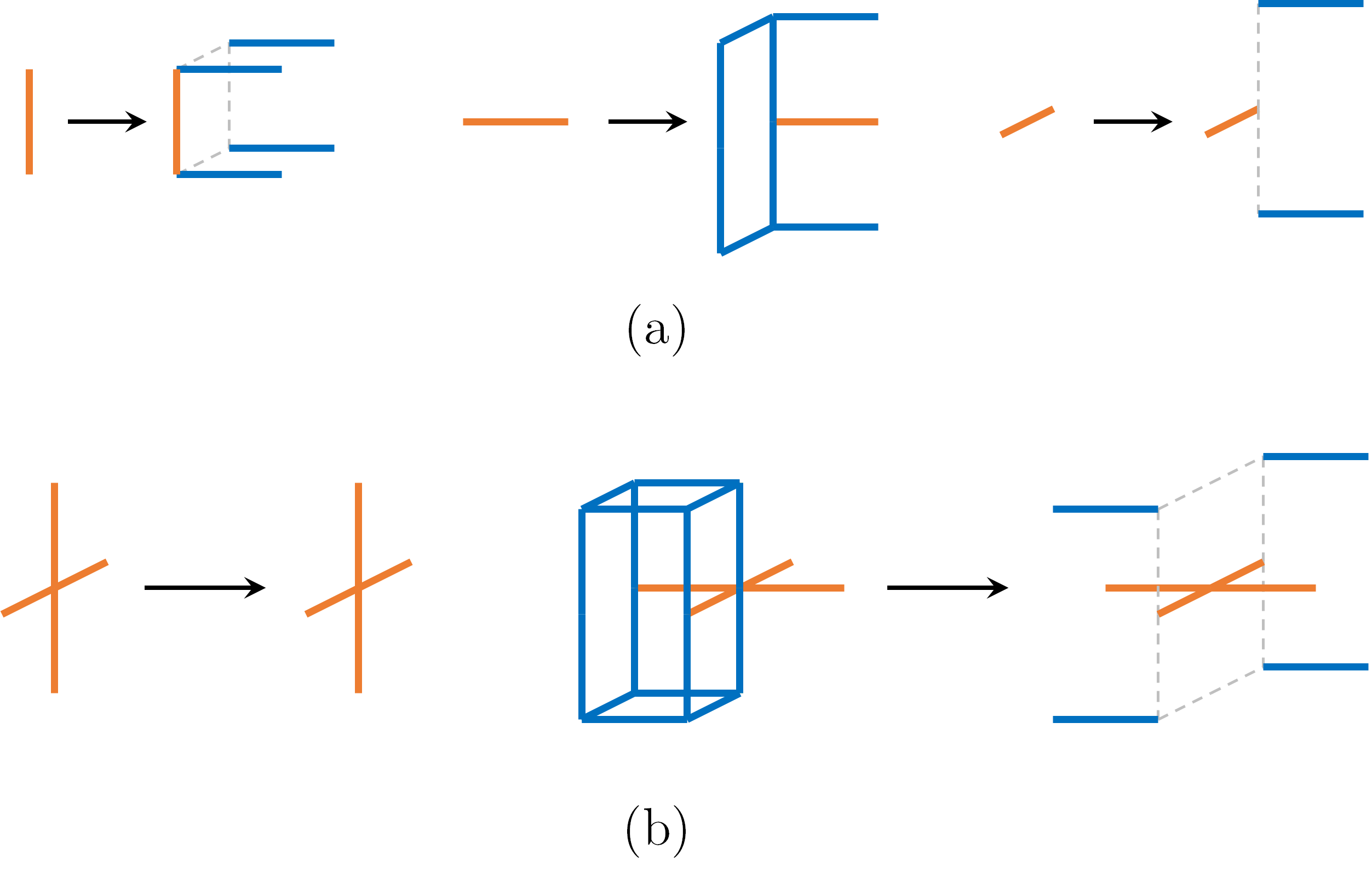}
    \caption{(a) Operator algebra isomorphism that maps the ground space of $H_\text{XC}$ to that of $H_\text{fXC}$. Here blue and orange represent Pauli $X$ and $Z$ respectively. The mapping leaves the Pauli $X$ operators unchanged. This transformation can be realize by a translation-invariant circuit of $HCZH$ gates. (c) The stabilizer generator $A_{(x-1/2,y-1/2,z-1/2)}A_{(x-1/2,y-1/2,z+1/2)}B^z_{(x,y,z)}$ of the stabilizer code defined by $H_\text{XC}$ is mapped to the flux constraint vertex term $\tilde{B}^z_{(x,y,z)}$ of $H_\text{fXC}$.}
    \label{fig:fXcto}
\end{figure}

There also exists a local unitary Clifford circuit, depicted in Fig. \ref{fig:fXcto}, that maps the ground space of the X-cube model to the ground space of the fermionic X-cube model, demonstrating that the two models have the same fractonic order. This circuit maps the stabilizer generator $A_{(x-1/2,y-1/2,z-1,2)}A_{(x-1/2,y-1/2,z+1/2})B_{(x,y,z)}^z$ of the $H_\text{XC}$ stabilizer code to the term $\tilde{B}_{(x,y,z)}^z$ of $H_\text{fXC}$. Here $A_c$ is the cube term and $B_v^z$ and $\tilde{B}_v^z$ are vertex terms. In similar fashion to the 2-foliated case, this mapping has the interpretation of flux attachment: in particular, a lineon dipole mobile in the $zx$ plane is attached to each fracton in the X-cube model. A natural conclusion is that this flux attachment transmutes the statistics of each fracton from bosonic to fermionic. It is an important question to establish an invariant meaning of fracton statistics to verify that this is the case.

We conjecture, but do not construct an explicit Clifford circuit, that the half-fermionic X-cube model $H_\text{hXC}$ is also equivalent to $H_\text{XC}$ via local unitary transformation. This conjecture is based on the observation that there is a similar flux attachment procedure that transmutes between the gauge charge statistics of the two models. In this case, lineon dipoles in all even planes in all three directions are attached to the fractons in their respective planes of mobility. Thus either 1 or 3 lineon dipoles are attached to each even checkerboard sublattice ($A$) fracton, and 0 or 2 to each odd sublattice ($B$) fracton. As a result the statistics of $A$ fractons are transmuted but not of $B$ fractons. Correspondingly all of the fracton dipole (planon charge) exchange statistics are transmuted from fermionic to bosonic.

\subsection{Fermionic Fibonacci prism model}

In the fermionic Fibonacci prism model, the fusion structures of gauge charge and gauge flux are inverse to one another with respect to spatial inversion. In particular, configurations of charge lineons at positions $i$, $i-\hat{x}+\hat{y}$, $i+\hat{y}$, and $i+\hat{x}+\hat{y}$ fuse to the vacuum sector, whereas configurations of flux lineons at positions $i$, $i+\hat{x}-\hat{y}$, $i-\hat{y}$, and $i-\hat{x}-\hat{y}$ likewise fuse to the vacuum. These configurations generate all possible configurations that fuse to the vacuum, hence there is no combination of fluxes which reproduces the fusion structure of a charge. As a result, there is no way to attach flux to a fermionic Fibonacci lineon charge to transmute its statistics while preserving its fusion rules. Therefore, the fermionic Fibonacci prism model must represent a novel fractonic order unique from its bosonic counterpart, distinguished by the exchange statistics of the gauge charge.

As an aside, we observe that it is also possible to view the fermionic Fibonacci prism model as a \textit{bosonic} gauge theory where flux in the fermionic gauge theory is reinterpreted as charge in the bosonic gauge theory. The subsystem symmetries giving rise to this model are Ising $Z_2$ symmetries acting on stacks of \textit{inverse} Fibonacci triangle fractals (there is one spin-1/2 degree of freedom per site of a cubic lattice). Since the gauge flux in the bosonic gauge theory has nontrivial, fermionic statistics, it is a \textit{twisted} fractal spin liquid in the sense that it is dual to a 3D fractal subsystem symmetry protected topological (SSPT) phase.\cite{YouSSPT,YouSondhiTwisted,DevakulSSPT,DevakulFractal} The Hamiltonian of this SSPT is
\begin{equation}
\begin{split}
    H=-\sum_i X_i &Z_{i-\hat{x}+\hat{y}-\hat{z}}Z_{i+\hat{y}-\hat{z}}Z_{i+\hat{x}+\hat{y}-\hat{z}}Z_{i-2\hat{x}-\hat{z}}\\
    \times&Z_{i+\hat{x}-\hat{y}+\hat{z}}Z_{i-\hat{y}+\hat{z}}Z_{i-\hat{x}-\hat{y}+\hat{z}}Z_{i+2\hat{x}+\hat{z}}.
\end{split}
\end{equation}

\section{Discussion}
\label{sec:discussion}

In this paper we have constructed a handful of novel exactly solvable models of fractonic order, by gauging \textit{subsystem fermion parity symmetries} via a general procedure. Systems with subsystem fermion parity symmetry are generally analogous to spin systems with subsystem Ising symmetry. Gauging the latter type of symmetry in trivial paramagnets is well-known to give rise to CSS stabilizer code models of fractonic order. We have seen that gauging the analogous subsystem fermion parity symmetry in trivial insulating states gives rise to non-CSS stabilizer code models of fractonic order with modified flux constraints. These models are characterized by emergent fermionic gauge theory, in which fractional excitations with constrained mobility may exhibit fermionic exchange statistics.

We have studied examples that are fermionic analogs of the X-cube model, the 2-foliated lineon model, and the Fibonacci prism model.\cite{YoshidaFractal} Remarkably, the analogs of the X-cube and 2-foliated models exhibit the same fractonic order as the bosonic counterparts obtained by gauging planar $Z_2$ symmetries. This result sheds a new light on the X-cube model: while it is well-known that its fractonic order can be viewed as an emergent bosonic gauge theory with subsystem gauge symmetry, these results show it is also possible to view it as an emergent fermionic gauge theory. A natural question to ask is whether there are planar fractonic orders that are characterized by strictly bosonic gauge theory or strictly fermionic gauge theory. It is likely that gauging planar subsystem symmetries in planar subsystem symmetry protected topological (SSPT) phases,\cite{YouSondhiTwisted,DevakulShirley,TwistedFoliated} whether fermionic or bosonic, is likely to yield models satisfying this criterion.

On the other hand, the fermionic Fibonacci prism model is a unique fractonic order from its bosonic counterpart, owing to the spatially inverted fusion structures of charge and flux. The gauging procedure can also be applied to subsystem fermion parity symmetries of other fractal geometries; for instance it can be used to generate a fermionic analog of the Haah code.\cite{HaahCode} It is likely that such models are also distinct from their bosonic counterparts for the same reason. These fermionic gauge theories also represent twisted versions of \textit{bosonic} gauge theories in which fermionic gauge charge is reinterpreted as gauge flux of the bosonic theory with nontrivial statistics. Ungauging these models gives rise to nontrivial three-dimensional fractal SSPT phases. A natural question to ask is whether there are other kinds of fractal SSPTs in 3D, and how to construct further models of this type and classify these phases.

The existence of fermionic planon gauge charge in the half-fermionic X-cube model raises another interesting possibility: the construction of chiral fracton models. This could be achieved by adding layers of 2D Kitaev 16-fold way states to the half-fermionic X-cube model and condensing bound states of 2D layer fermionic gauge charge and fracton dipole fermionic gauge charge of the half-fermionic X-cube model. It may be interesting to study the resulting fractonic orders in detail, especially in terms of potential foliation structure.

Finally, we note that the gauging procedure we have introduced strongly suggests that gauge charge excitations in certain models, such as the fermionic X-cube model, are simultaneously fermions and fractons. We have argued that because the fermionic X-cube fractons become fermionic lineons upon condensation of a stack of fracton dipoles, they are themselves fermions. These observations suggest two natural directions of further study which will be addressed in future work. First, it would be interesting to establish a general theory of boson condensation in gapped fracton phases, analogous to the theory of boson condensation in 2D topological orders,\cite{Fiona,BoseCondensation2,BoseCondensation1} which would solidify this line of reasoning. Second, it would be worthwhile to study how the fermionicity of fracton excitations manifests via long-range statistical interactions. In particular, are there processes that define an invariant notion of fracton `exchange' statistics? This question is pertinent not only to planar fracton orders such as the fermionic X-cube model but also fractal spin liquids such as the fermionic Haah code.

Near the completion of this manuscript, we became aware of similar results, obtained independently in Ref. \onlinecite{Nat}.

\acknowledgements

We are grateful to Xie Chen, Michael Hermele, Kevin Slagle, and Nathanan Tantivasadakarn for helpful discussions. The author is supported by the National Science Foundation under award number DMR-1654340 and the Institute for Quantum Information and Matter at Caltech, and by the Simons collaboration on ``Ultra-Quantum Matter''.

\appendix
\section{X-cube model from gauging planar symmetry}
\label{app:Xcube}

The X-cube model is obtained by gauging the intersecting $xy$, $yz$, and $zx$ planar symmetries of the trivial paramagnetic Hamiltonian with one spin-1/2 degree of freedom at each site of a cubic lattice:
\begin{equation}
    H=-\sum_iX_i.
\end{equation}
The $Z_2$ planar symmetry for plane $P$ takes the form
\begin{align}
    S_P=\prod_{i\in P}X_i.
\end{align}
The algebra of subsystem symmetric observables is generated by the on-site symmetry representations $X_i$ and the minimal coupling terms, which are products of $Z$ operators around the four vertices of each plaquette.
To gauge the symmetries, $Z_2$ gauge fields are placed on each elementary plaquette $p$, and generalized Gauss's laws
\begin{equation}
    X_i\prod_{p\ni i}X_p=1
\end{equation}
are imposed as constraints on the expanded Hilbert space. Finally, local terms commuting with the Gauss's law constraints are added to the Hamiltonian to gap out gauge flux excitations. These terms act on the gauge degrees of freedom and take the form
\begin{equation}
    A^z_c=\prod_{p\in c_{xy}}Z_p,
\end{equation}
where $c_{xy}$ denotes the 4 faces of cube $c$ normal to the $xy$ plane. There are three such constraints, $A_c^x$, $A_c^y$, and $A_c^z$, for each elementary cube in the lattice. To connect with the usual representation of the X-cube model, we suppress the matter degrees of freedom by mapping the constraint $X_i\prod_{p\ni i}X_p=1$ to $X_i=1$ via the change of variables given by the local unitary
\begin{equation}
    \prod_i\left(\prod_{p\ni i}C_vX_p\right),
\end{equation}
where $C_vX_p$ denotes the two-qubit controlled-$X$ gate with control qubit the matter spin at vertex $v$, and target qubit the gauge field on plaquette $p$. 
Finally, we switch from the direct lattice to the dual lattice, so that the gauge fields now reside on edges $e$ and the flux constraints are mapped into the vertex constraints $A_v^x$, $A_v^y$, and $A_v^z$. In the constrained subspace, the resulting Hamiltonian takes the form
\begin{equation}
    H_\text{XC} = -\sum_c\prod_{e\in c}X_e -\sum_v (A_v^x+A_v^y+A_v^z).
\end{equation}

\section{Fracton dipole condensation in the fermionic X-cube model}
\label{app:condensation}

\begin{figure}
    \centering
    \includegraphics[width=0.45\textwidth]{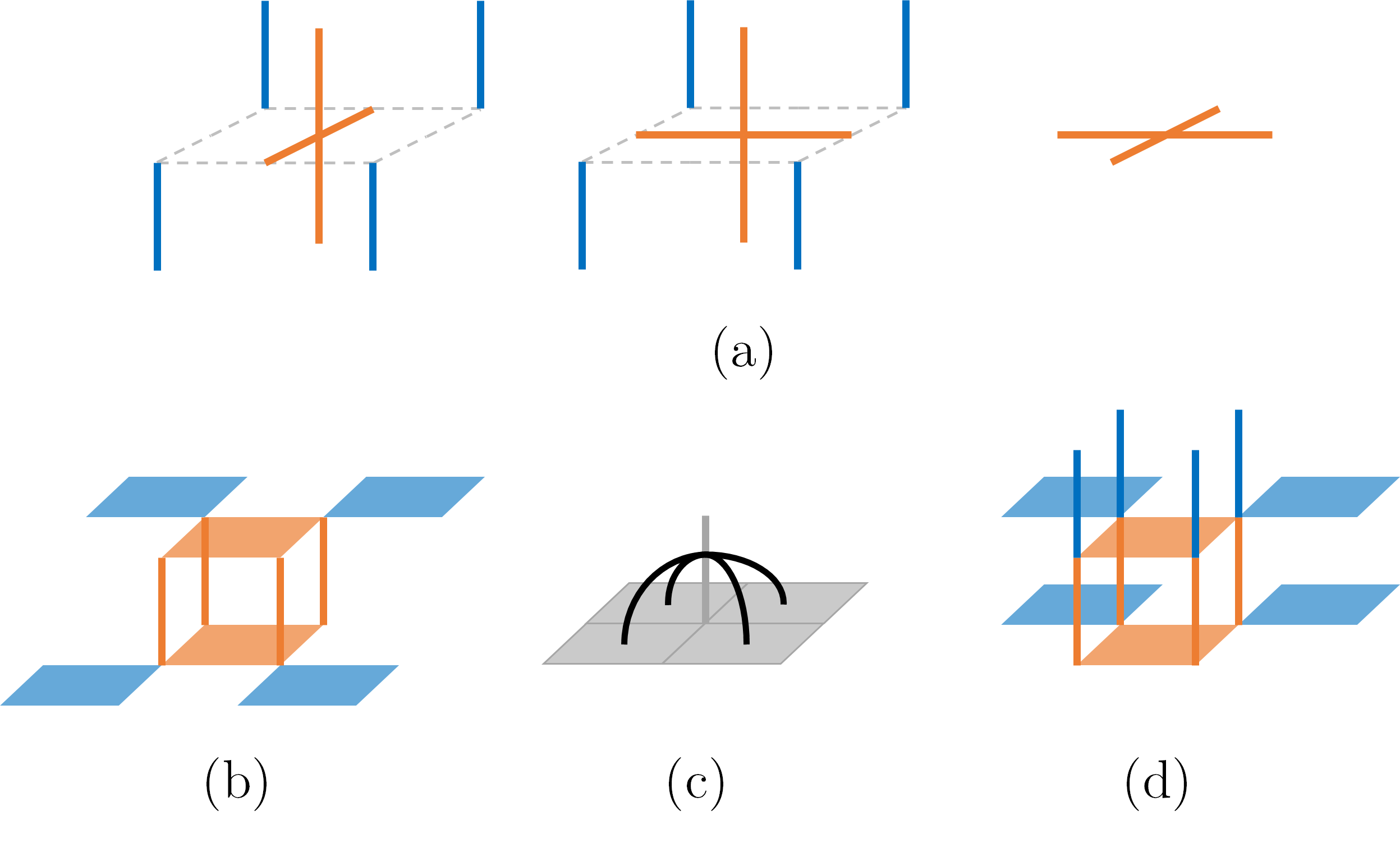}
    \caption{(a) Rotated versions $\bar{B}_c^\mu$ of the terms $\tilde{B}_c^\mu$. (b) The term $\tilde{D}_c$ of $H_\text{condensed}$. (c) Local unitary transformation that transforms $\tilde{D}_c$ into $\tilde{B}_c$ of $H_\text{f2F}$ after augmenting with an additional $Y\leftrightarrow Z$ unitary on each qubit. A unit cell of the circuit is depicted, composed of four two-qubit $HCZH$ gates represented by the black arcs. Here $H$ is Hadamard rotation and $CZ$ the controlled-$Z$ gate. (d) The term $\tilde{B}_c$ of $H_\text{f2F}$.}
    \label{fig:condense}
\end{figure}

In this section we will describe a procedure in which a stack of fracton dipoles in the $z$ direction are condensed in the fermionic X-cube model $H_\text{fXC}$, driving a phase transition into a new phase whose fixed point is the fermionic 2-foliated lineon code model $H_\text{f2F}$. As a result of the condensation, all of the fracton superselection sectors along a line parallel to the $z$ axis collapse to a single lineonic superselection sector, which corresponds to the fermionic lineon gauge charge excitation of $H_\text{f2F}$. This mechanism is similar to the construction of the 3D toric code from layers of 2D toric codes, in which pairs of 2D $e$ particles are condensed to give rise to a single mobile 3D particle.\cite{QiJian}

To describe this condensation transition, we will actually begin with a model $H_\text{fXC}'$ that is equivalent to but slightly different from $H_\text{fXC}$. In the direct lattice picture, the Hilbert space of $H_\text{fXC}$ consists of one qubit on each plaquette of a cubic lattice, and has the form
\begin{equation}
    H_\text{fXC}=-\sum_vA_v -\sum_v \left(\tilde{B}_c^x+\tilde{B}_c^y+\tilde{B}_c^z\right)
\end{equation}
(Here $A_v$ and $\tilde{B}^\mu_c$ correspond to $A_c$ and $\tilde{B}^\mu_v$ on the dual lattice.)
The modified system has an additional qubit on each $z$-oriented link $l_z$ of the direct lattice, and the Hamiltonian has the form
\begin{equation}
    \begin{split}
    H'_\text{fXC}&=-\sum_vA_v -\sum_{l_z}A_{l_z} -\sum_v \left(\tilde{C}_c^x+\tilde{C}_c^y+\tilde{C}_c^z\right)\\
    A_{l_z}&=X_{l_z}\prod_{p\ni l_z}X_p\qquad\qquad
    \tilde{C}_c^z=\bar{B}_c^z\\
    \tilde{C}_c^x&=\bar{B}_c^x\prod_{l_z\in c}Z_{l_z}\qquad\qquad
    \tilde{C}_c^y=\bar{B}_c^y\prod_{l_z\in c}Z_{l_z}
    \end{split}
\end{equation}

Here, $\bar{B}_c^\mu$ are rotated versions of $\tilde{B}_c^\mu$ depicted in Fig. \ref{fig:condense}a. To condense all fracton dipoles mobile in the $xy$ plane, the terms $Z_p$ are added to the Hamiltonian for plaquettes $p$ normal to the $x$ or $y$ direction:
\begin{equation}
    H=H'_\text{fXC}-J\sum_{p\perp \hat{x},\hat{y}}Z_p
\end{equation}
and the limit $J\to\infty$ is taken, freezing the $x$-normal and $y$-normal plaquette degrees of freedom in the $Z_p=1$ state. The effective Hamiltonian on the remaining Hilbert space takes the form
\begin{equation}
    H_\text{condensed}=-\sum_vA_v^\text{2F}-\sum_c\tilde{D}_c
\end{equation}
where $A_v^\text{2F}$ is equal to the vertex term of $H_\text{2F}$ or $H_\text{f2F}$, and emerges as the product of $H'_\text{fXC}$ terms $A_v\prod_{l_z\ni v}A_{l_z}$. Thus excitations of $A_v$ prior to condensation correspond to excitations of $A_v^\text{2F}$ after condensation. The cube term $\tilde{D}_c$ is depicted in Fig. \ref{fig:condense}b, and is mapped to the cube term $\tilde{B}_c$ of $H_\text{f2F}$ via the local unitary transformation of Fig. \ref{fig:condense}c, which leaves $A_v^\text{2F}$ invariant. This demonstrates that $H_\text{condensed}$ describes the same phase as $H_\text{f2F}$, and that the fractonic gauge charges of $H_\text{fXC}$ are mapped to the fermionic lineon charges of $H_\text{f2F}$ across the condensation transition. This suggests that the fractons of the fermionic X-cube model are themselves fermions.

\section{Half-fermionic X-cube model}
\label{app:halfXcube}

\begin{figure}
    \centering
    \includegraphics[width=0.47\textwidth]{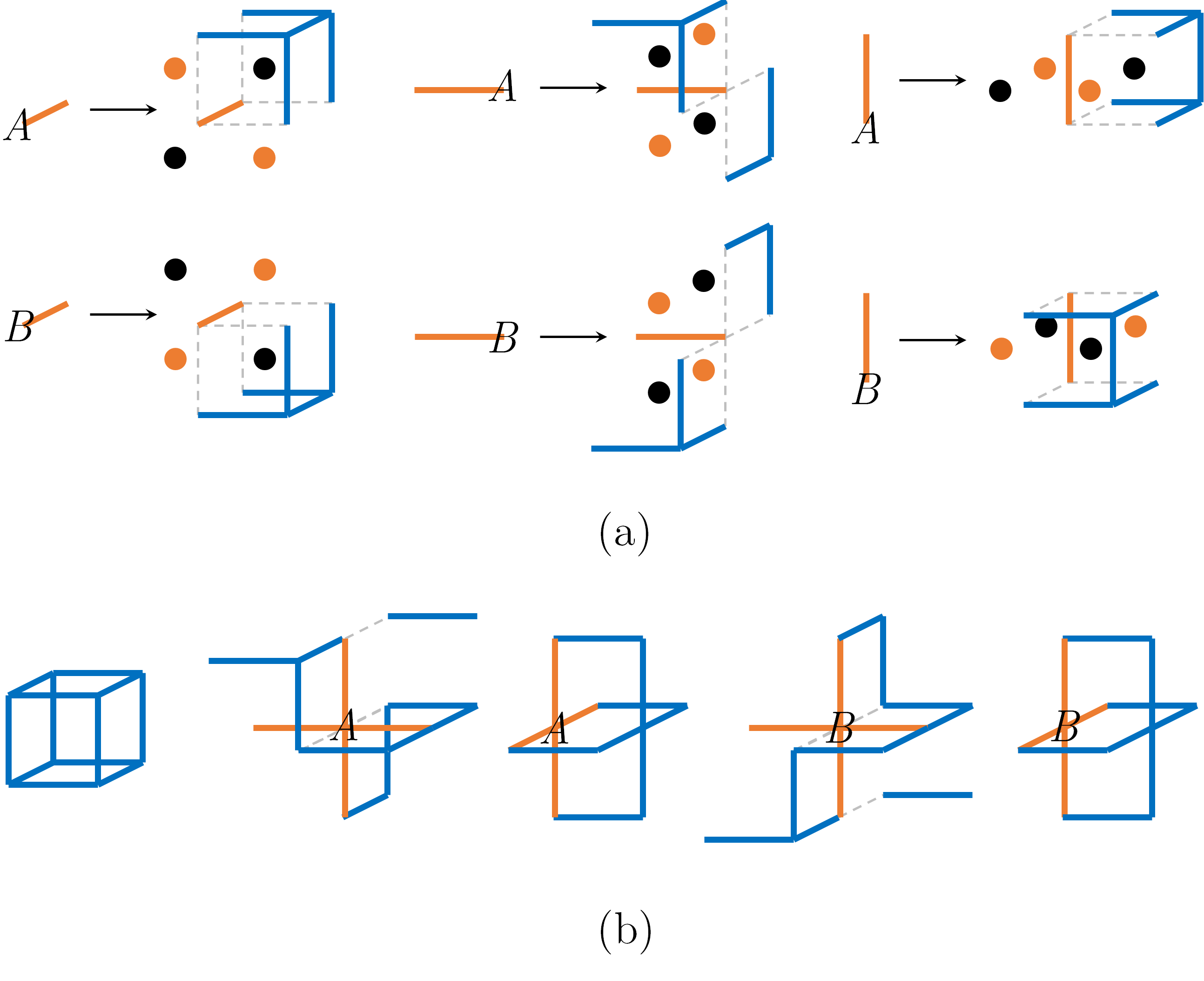}
    \caption{(a) Mapping of operators $Z_p$ in the operator algebra isomorphism. This mapping has the translation symmetry of the checkerboard sublattice. The factors of $X_p$ are added to ensure that the commutation of $Z_p$ and $Z_q$ are preserved. Here we have switched to the dual lattice picture in which gauge qubits lie on edges. Blue represents Pauli $X$, orange represents Pauli $Z$, and the black dots represent $\gamma$. (b) Cube and vertex terms of the Hamiltonian $H_\text{hXC}$. The form of the flux constraints differ between the $A$ and $B$ sublattices. The third vertex term on each sublattice is generated by the other two.}
    \label{fig:hXC}
\end{figure}

In this appendix we describe the half-fermionic X-cube model. As discussed in Section \ref{sec:half}, the following Hamiltonian is obtained by gauging planar subsystem symmetries in a cubic lattice system with one fermionic mode on every $A$ sublattice site and one spin-1/2 degree of freedom on every $B$ sublattice site:
\begin{equation}
    {H}_g=i\sum_{i\in A}\gamma_i\gamma'_i-\sum_{i\in B}X_i -\sum_c \left(B_c^x+B_c^y+B_c^z\right)
\end{equation}
This Hamiltonian can be transformed so it acts on a purely bosonic tensor product Hilbert space, via the following operator algebra isomorphism generalizing Eq. \ref{eq:transform}:
 \begin{equation}
        \begin{split}
            \gamma_i\to \gamma_i\prod_{p\ni i} X_p\qquad \gamma_i'\to\gamma_i'\qquad &i\in A
            \\
            X_i\to X_i\prod_{p\ni i} X_p\qquad Z_i\to Z_i\qquad &i\in B
        \end{split}
\end{equation}
Furthermore the transformation maps $X_p\to X_p$ and $Z_p$ as depicted in Fig. \ref{fig:hXC}a. Here $p$ refers to plaquettes of the original lattice. This isomorphism maps the Gauss's law constraints to the on-site constraints $-i\gamma_i\gamma'_i=1$ or $X_i=1$, thus suppressing the matter degrees of freedom. The resulting spin Hamiltonian $H_\text{hXC}$ is depicted in Fig. \ref{fig:hXC}b.

\end{document}